\documentclass[useAMS,usenatbib]{mn2e}
\usepackage{bm}
\usepackage{graphicx}
\usepackage{fixltx2e}
\newcounter{ionstage}
\newcommand{\ion}[2]{\setcounter{ionstage}{#2}%
  \ensuremath{\mathrm{#1\,\scriptstyle\Roman{ionstage}}}}
\newcommand\nii{[\ion{N}{2}]}
\newcommand\sii{[\ion{S}{2}]}
\newcommand\oiii{[\ion{O}{3}]}
\newcommand\ha{\ensuremath{\mathrm{H\alpha}}}
\newcommand\hii{\ion{H}{2}}

\newcommand\gammaVCAthin{\ensuremath{\gamma_{\mathrm{t}}}}
\newcommand\gammaVCAvthick{\ensuremath{\gamma_{\mathrm{T}}}}
\newcommand\mSF{\ensuremath{m_{\mathrm{2D}}}}

\nonstopmode

\title[Turbulence in Simulated HII regions]{Turbulence in simulated HII regions}
\author[S.-N. X. Medina et al.]{S.-N. X. Medina$^{1,2}$, S. J. Arthur$^{2,4}$\thanks{E-mail:
j.arthur@crya.unam.mx}, W. J. Henney$^{2,4}$, G. Mellema$^{3,4}$, A. Gazol$^{2}$\\
$^{1}$Departamento de Astronom\'{\i}a, Universidad de Guanajuato,
Guanajuato, M\'exico.\\
$^{2}$Centro de Radioastronom\'{\i}a y Astrof\'{\i}sica, Universidad
Nacional Aut\'onoma de M\'exico, Campus Morelia, 58090 Morelia,
Michoac\'an, M\'exico.\\
$^3$Dept.\ of Astronomy and Oskar Klein Centre, AlbaNova, Stockholm
University, SE-106 91 Stockholm, Sweden.\\
$^4$Nordita, KTH Royal Institute of Technology and Stockholm
University, Roslagstullsbacken 23, SE-106 91 Stockholm, Sweden.}
\begin{document}

\date{}

\pagerange{\pageref{firstpage}--\pageref{lastpage}} \pubyear{2013}

\maketitle

\label{firstpage}

\begin{abstract}
We investigate the scale dependence of fluctuations inside a realistic
model of an evolving turbulent HII region and to what extent these
may be studied observationally.  We find that the multiple scales of
energy injection from champagne flows and the photoionization of
clumps and filaments leads to a flatter spectrum of fluctuations than
would be expected from top-down turbulence driven at the largest
scales.  The traditional structure function approach to the
observational study of velocity fluctuations is shown to be
incapable of reliably determining the velocity power spectrum of our
simulation.  We find that a more promising approach is the Velocity
Channel Analysis technique of Lazarian \& Pogosyan (2000), which,
despite being intrinsically limited by thermal broadening, can
successfully recover the logarithmic slope of the velocity power
spectrum to a precision of \(\pm 0.1\) from high resolution optical
emission line spectroscopy.
\end{abstract}

\begin{keywords}
hydrodynamics --- HII regions --- ISM: kinematics and dynamics --- turbulence
\end{keywords}

\section{Introduction}

The line broadening in excess of thermal broadening seen in optical
spectroscopic studies of \hii{} regions has been attributed to
turbulence in the photoionized gas. There are many examples in the
literature of attempts to identify the presence and characterize this
turbulence, for example \citet{1958RvMP...30.1035M}, \citet
{1985ApJ...288..142R}, \citet
{1987ApJ...317..686O}, \citet{1995ApJ...454..316M}, \citet{1997ApJ...487..163M}, 
 \citet{2011MNRAS.413..721L} and references cited by these papers. In
these studies, the variation of the point-to-point radial velocities
with scale was investigated using structure functions following  \citet{{1951ZA.....30...17V}}.

A general finding of these studies is that the structure function
derived from these observations does not follow that predicted by
Kolmogorov's law \citep {1941DoSSR..30..301K} . The interpretation of this result is that
Kolmogorov's law for incompressible and subsonic turbulent flows
cannot be strictly applied to the photoionized gas in \hii{}
regions, which will be compressible and possibly mildly supersonic
\citep{1985ApJ...288..142R}, or that energy does not cascade from
larger to smaller scales but is input at many scales
\citep{{1988ApJS...67...93C}}. Moreover, the results differ according
to which emission lines are used in the study, since the topology of
the emitting gas will be different. For instance,
the \oiii$\lambda$5007 line comes from gas in the interior of an
\hii{} region, whereas the \sii$\lambda$6731 line comes from the
vicinity of the ionization front, essentially a two-dimensional
surface. The suggested mechanisms responsible for the generation and
maintenance of the turbulent velocities in \hii{} regions include
photoevaporated flows from globules
\citep{{1968Ap&SS...1..388D},{1987ApJ...317..686O}} and stellar winds
\citep{{1997ApJ...487..163M},{2011MNRAS.413..721L}}.


In order to construct the structure functions for the velocity fields,
observations at many points in an \hii{} region are needed. This
can be achieved either by multiple longslit spectroscopic observations at
many positions across a nebula
\citep{{1987ApJ...317..676O},{1987ApJ...315L..55C},{1993ApJ...409..262W}}, or by Fabry-Perot
interferometry
\citep{{1985ApJ...288..142R},{1995ApJ...454..316M},{2011MNRAS.413..705L}}. Longslit observations with high velocity resolution have enabled several
velocity components to be identified for emission lines of metal ions
for which the thermal widths are small, e.g., \oiii$\lambda$5007. These
observations have been used to determine the radial velocities of the
principal components of the emitting gas at hundreds of positions
within the nebula. Fabry-Perot observations produce datasets of
thousands of radial velocities, but without the velocity resolution to
distinguish between different velocity components. Obviously, in order
to obtain a structure function over a wide range of scales, very high
quality data with an ample spatial coverage are required. In the case
of longslit spectra, this is a non-trivial task, not least the
calibrating of the positions of the slits
\citep{{2007AJ....133..952G},{2008RMxAA..44..181G}}.


\citet{1949ApJ...110..329C} suggested that turbulence is ubiquitous in
astrophysics and that it could be analysed statistically via the
spectrum of the velocity field, which expresses the correlations
between instantaneous velocity components at all possible pairs of
points in the medium. 
Turbulence develops in a fluid when velocity advection is dominant
over dissipation (high Reynolds number), then
continuous injection of energy is required in order to maintain the turbulent state. One of the main
assumptions of the Kolmogorov theory of turbulence is that energy is only injected at large scales
and is dissipated only at small scales. This implies that the energy cascades from large to small
scales without dissipation. Another assumption of this theory is that in the range between
injection and dissipation scales the energy is transferred at a
constant rate, and this range is
called the inertial range.
Turbulence is often described in terms of the energy
spectrum $E(v)$, and the inertial range is represented by a power-law
relationship $E(v) \propto k^\beta$, where $k$ is the wave number.
For incompressible, homogeneous, isotropic, 3D turbulence \citep{1941DoSSR..30..301K},
we have $\beta = -5/3$, while in the limit of high Mach number,
shock-dominated turbulence in one dimension \citep{1974Burgers}, the
power law is $\beta = -2$. It has proved more difficult to obtain
exact results for the general case of 3D compressible, hydrodynamic
turbulence. Scaling laws suggest that the original Kolmogorov energy
spectrum scaling should be preserved even for highly compressible
turbulence if the density weighted velocity $\rho^{1/3}v$ is taken
instead of the pure velocity $v$ \citep{{2007ApJ...665..416K},
  {2010A&A...512A..81F}}. Recent work by \citet{2011PhRvL.107m4501G}
predicts the relation $E(\rho^{1/3}v) \propto k^{-19/9}$ for
compressible, isothermal turbulence with compressive driving, with a
turnover around the sonic scale to $E(\rho^{1/3}v)\propto
k^{-5/3}$. This has been confirmed by extremely high resolution
numerical experiments \citep{2013MNRAS.436.1245F}.

Another important prediction of the Kolmogorov theory is the scaling of the structure function of
any order. The universality of such scaling laws has been tested by detailed
numerical experiments
\citep{{2002PhRvE..66b6301P},{2002ApJ...573..678B},{2007ApJ...665..416K},{2013MNRAS.436.1245F}}. 
Even in numerical experiments of incompressible, homogeneous and isotropic turbulence with
energy injection at a fixed large scale, it is found that the exponents of the higher order structure
functions have an anomalous behaviour, which has been interpreted as the result of spatial and/or
temporal intermittency.
Intermittency is a sparseness in space and time
of strong structures associated with the dissipation or injection of
energy \citep{1994PhRvL..72..336S}. Since astrophysical turbulence driven by real
physical phenomena is almost certainly intermittent, it is important
to understand the possible effects of intermittency on the derived
statistical properties of real \hii{} regions.


The three-dimensional properties of the density and velocity fields
that describe the turbulence in the interstellar medium, in this
case an \hii{} region, must be deduced from the statistical
properties of the observed quantities. For photoionized gas, the
observations generally consist of spectra at different positions, from
which centroid velocities and linewidths can be obtained. For optical
line emission, the emissivity depends on density, temperature and
ionization state and therefore an analysis of the emission lines will
not just be providing information on the turbulent velocity
fluctuations in the gas. The observations are a two-dimensional
projection of the three-dimensional properties and 
much effort has been dedicated to the problem of recovering 3D
information from 2D data
\citep{{2000ApJ...537..720L},{2002ApJ...566..276B},{2003ApJ...593..831M},
  {2004ApJ...604..196B},{2005ApJ...631..320E},{2010MNRAS.405L..56B}}. Generally,
some simplifying assumptions, such as isothermal gas and statistical isotropy of the
turbulence, must be made.


Recent modeling of \hii{} regions has addressed the origin of the
irregular structures, filaments, and globules seen within and around
the borders in optical images \citep{{1996ApJ...469..171G},
  {2001MNRAS.327..788W},{2006ApJ...647..397M},
  {2010MNRAS.403..714M},{2011MNRAS.414.1747A}}. Morphologically, the
simulated emission-line images are very reminiscent of observed \hii{}
regions and the global dynamics, as measured by the r.m.s.\ velocity,
is also similar to observationally derived values
\citep{{2008RMxAA..44..181G},{2011MNRAS.414.1747A},{2011MNRAS.413..705L}}. The
internal dynamics of the simulated \hii{} regions is due mainly to the
interaction of photoevaporated flows from the heads of the filaments
and clumps, which flow into the interior of the \hii{} region,
superimposed on the general expansion of the \hii{} region
\citep{{2006ApJ...647..397M}, {2011MNRAS.413..705L}}. However,
the dynamics of real \hii{} regions could also be affected at large
scales by the action of stellar winds from the ionizing star or stars,
and at small scales by outflows from young, low-mass stars.


In this paper we investigate to what extent our numerical simulations of
\hii{} regions model observed statistical properties of real \hii{}
regions. In \S~\ref{sec:nummod} we briefly describe the numerical
methods used in the radiation-hydrodynamics simulations and in the
calculation of the simulated emission-line radiation, and how the
statistical information is obtained from these calculations. In
\S~\ref{sec:results} we describe our results for the expansion of an
\hii{} region in a turbulent, clumpy molecular cloud. We discuss these results in
\S~\ref{sec:discuss}, and comment on the extent to which the
statistical properties of  the numerical simulations agree with those
of observed \hii{} regions. In \S~\ref{sec:summary} we summarize and present our conclusions.

\section[]{Numerical Model}
\label{sec:nummod}
We perform radiation-hydrodynamic simulations of the expansion of
photoionized regions in a non-uniform initially neutral
medium. The central ionizing source has an effective temperature of
37,500~K and an ionizing photon rate of $10^{48.5}$~s$^{-1}$, which
corresponds approximately to an O7 star.

The simulation takes as an initial condition a clumpy medium with mean
density $\langle n_0 \rangle= 1000$~cm$^{-3}$, which results from a
self-gravitating forced turbulence simulation by
\citet {2005ApJ...630L..49V}. The computational cube represents a
spatial volume of
4~pc$^3$ and the initial temperature is 5~K. 

The radiation-hydrodynamics code used in the present paper does not
include self gravity. However, although the initial conditions would
be gravitationally unstable in the absence of photoionization heating,
on the relatively short timescales of
the simulations ($\sim 3\times10^5$~yrs) this will not be important
since the global free-fall time of the computational cube is $\sim
6\times10^5$~yrs. 

\subsection{Radiation-hydrodynamics Code}
\label{subsec:rhdcode}
The radiation-hydrodynamics code used in this paper is the same as
that used by \citet{2006ApJ...647..397M}. The hydrodynamics is calculated
using the nonrelativistic Roe solver PPM scheme described in
\citet{1995A&AS..110..587E} with the addition of a local oscillation filter
\citep{{2003ApJS..147..187S}} to suppress numerical odd-even decoupling
behind radiatively cooling shock waves. Heating and cooling in the
ionized and neutral gas is dealt with in the same manner as described
in detail by \citet{2009MNRAS.398..157H}. The radiation transport and
photoionization makes use of the C$^2$-RAY (Conservative-Causal ray)
code developed by \citet {2006NewA...11..374M}. The calculations are
all performed on a fixed, uniform Cartesian grid in three dimensions
with a resolution of $512^3$ cells.

\subsection{Simulated Emission Lines}
\label{subsec:simemiss}
The emissivity cubes for the H$\alpha$ recombination line and the
\oiii$\lambda$5007, \nii$\lambda$6584 and \sii$\lambda$6731
collisionally excited lines are calculated as described in
\citet{{2011MNRAS.414.1747A}}, with the assumption that the heavy
element ionization fractions are fixed functions of the hydrogen
ionization fractions, calibrated with the Cloudy photoionization code
\citep{2013RMxAA..49..137F}. 

\subsection{Turbulence Statistics}
\label{subsec:turbstats}
The emissivity statistics have contributions from velocity, density,
temperature and ionization state fluctuations. Obviously, the velocity
fluctuations are the most dynamically important statistic and spectral
line data, both real and simulated, can be used to extract statistical
information for the different contributions. In particular, the
second-order structure function of the velocity centroids and the
technique of velocity-channel analysis (VCA; \citealp{2000ApJ...537..720L}) have been widely used.

In this paper, we calculate the second-order structure function
of the velocity centroids of the simulated emission lines and also apply
VCA to the corresponding position-position-velocity (PPV)
datacubes. In addition, we calculate the power spectrum of the velocity field,
both weighted and unweighted, of the 3D hydrodynamic simulation
datacube in an attempt to relate the observed statistics to the
underlying hydrodynamics.

\subsubsection{Power Spectrum}

The power spectrum is a statistical tool that is useful for describing
the intrinsic properties of, for example, velocity and density fields
or any other physical property. It is the Fourier
transform of the auto-correlation function of the physical
quantity. For example, the N-dimensional auto-correlation function of the
physical quantity $a$ can be written
\begin{equation}
  \xi_N(\boldsymbol{l}) = \langle
  a(\mathbf{r})a(\mathbf{r}+\boldsymbol{l})\rangle \ ,
\label{eq:corrfn}
\end{equation}
where $\mathbf{r}$ is the spatial position and $\boldsymbol{l}$ is the
spatial separation. The N-dimensional power spectrum is then
\begin{equation}
  p_N(k) = \int e^{i\mathbf{k}\cdot\boldsymbol{l}} \xi_N({\boldsymbol{l}}) \,
  d\boldsymbol{l} \ ,
\end{equation}
where $\mathbf{k}  = (k_x,k_y,k_z)$ is the wavenumber, which is related to the scale by
$k = 2\pi/l$, and the integration is performed over all N-dimensional
space. The energy spectrum is the angle integral of
the power spectrum over shells of radius $k = |\mathbf{k} |$, such that $E_N(k) \propto k^{N-1} p_N(k)$.

Often the power spectrum can be represented by a power law $p_3(k)
\propto k^{n}$. For instance, the velocity field for incompressible,
homogeneous \citep {1941DoSSR..30..301K}  turbulence has $n = -11/3$,
while for  shock-dominated  turbulence \citep{1974Burgers}, the 3D
power-law index is $n = -4$.

\subsubsection{Velocity centroid statistics}
\label{subsubsec:centroid}
The use of positional fluctuations in the velocity centroids of
spectral lines as probes of turbulent gas motions was developed in the
1950s by, for example, \citet {1958RvMP...30.1035M}, and has been
applied to observations of molecular emission lines in molecular cloud
complexes \citep{{1985ApJ...295..479D},{1985ApJ...295..466K}} and to optical emission lines
in the Orion Nebula, galactic and extragalactic \hii{} regions
\citep{{1985ApJ...288..142R},{1987ApJ...317..686O},{1988ApJS...67...93C},
  {2011MNRAS.413..705L}}. 

The plane-of-the-sky velocity centroid map is calculated from the
first two velocity moments $M_j$ of the simulated line intensity maps
$I(v_r)$, where
\begin{equation}
  M_j = \int_{v_{s1}}^{v_{s2}} v_s^j I(v_s) dv_s \ ,
\end{equation}
and $v_s$ is the radial (line-of-sight) velocity, for convenience taken to be
along one of the principal axes of the 3D data cube. 
The centroids are then
$V_c(x,y) = M_1/M_0$, where $(x,y)$ is the projection plane (plane of
the sky) when the line of sight is along the $z$-axis. The limits of the integration over velocity are 
$v_{s1}$ and $v_{s2}$, which represent the full range of
velocities produced by the simulations plus thermal broadening, where the thermal
width is $v_T = (kT/m_p)^{1/2}$ and $m_p$ is the atomic mass. 

The observed second-order structure function is
\begin{equation}
\label{eq:strucfunc}
  S_2(\boldsymbol{l}) = \frac{\Sigma[V_c(\mathbf{r}) - V_c(\mathbf{r} +
    \boldsymbol{l})]^2}{\sigma_c^2 N(\boldsymbol{l})} \ ,
\end{equation}
where the variance of centroid velocity fluctuations is
\begin{equation}
\label{eq:variance}
\sigma^2_c \equiv \frac{\Sigma [V_c(\mathbf{r}) - \langle V_c \rangle
  ]^2}{N} \ ,
\end{equation}
and $\langle V_c \rangle$ is the mean centroid velocity 
\begin{equation}
\label{eq:mean}
  \langle V_c \rangle \equiv \frac{\Sigma V_c(\mathbf{r})}{N} \ . 
\end{equation}
In this definition, $\mathbf{r}$ is the
two-dimensional position vector in the plane of the sky and
$\boldsymbol{l}$ is the lag, or separation vector. The summation in
Equation~\ref{eq:strucfunc} is over all data pairs for each
separation, $N(\boldsymbol{l})$, while the summations in the centroid
variation and mean are over the total number of array elements, i.e.,
pixels in the $(x,y)$-plane.

For homogeneous, incompressible (Kolmogorov) turbulence, the velocity fluctuations
scale as $l^{1/3}$ and hence the second-order structure function
scales like $l^{2/3}$. For isotropic velocity fields,  the structure function and the
auto-correlation function (see Eq.~\ref{eq:corrfn}) have the same scaling
and differ only by a constant. The structure function is therefore
related to the power spectrum. If the power-law index of the 3D structure
function is $m$, then the power-law index of the 3D power spectrum is
$n = -3 -m$.

\subsubsection{Velocity channel analysis}
\label{sec:stats-vca} 
The relationship between the velocity centroids  and the statistics of
the velocity field is only reliable if the density fluctuations are
negligible. Velocity channel analysis was developed to extract the
separate contributions of density and velocity from spectral line data cubes.

This is a technique for analyzing position-position-velocity (PPV) cubes
developed by \citet{2000ApJ...537..720L}. With this method, spectroscopic observations
are not reduced to velocity centroids as a function of position on the
plane of the sky. Instead, the PPV cubes are analyzed in terms of
velocity channels, or slices, as a function of the velocity resolution
used. As the width of the velocity slices increases, the relative
contribution of a velocity fluctuation to the total intensity
fluctuations decreases, because the contributions from many velocity
fluctuations will be averaged out in thicker velocity
slices. A slice is described as \textit{thick} when the dispersion of
turbulent velocities is less than the velocity slice thickness on the
turbulence scale studied, otherwise the slice is \textit{thin}
\citep{{2003MNRAS.342..325E}}. In the thickest velocity channels, we
obtain only information about the density fluctuations, since the
velocity information is averaged out. Conversely, the velocity
fluctuations dominate in thin channels.

Velocity channel analysis (VCA) consists of obtaining the 2D power
spectrum for each velocity channel and then averaging over all
velocity channels for each PPV cube. \citet{2003MNRAS.342..325E}
stress that whether a slice in velocity space is considered
\textit{thin} or \textit{thick} depends not only on the slice width
$\delta v = (v_\mathrm{max} - v_\mathrm{min})/N$, where $N$ is the
number of channels, but also on the scale of the turbulence. For
power-law statistics, the velocity dispersion scales as $\sigma_r
\propto r^{m/2}$, where $m$ is the velocity structure function
index. The criterion for a channel to be considered thin is $\delta v
< \sigma_r$, hence we do not expect a pure power-law result from the
averaged 2D power-spectra analysis. This is because the largest
scales, of the order of the size of the cube, are almost always in the
\textit{thin} regime. At the smallest scales, numerical dissipation
plays a r\^ole, and there is an additional deviation from the  power
law. 

Taking into account the thermal width of the optically thin emission
line introduces an additional limitation on the resolution that can be
used to discriminate between the \textit{thick} and \textit{thin}
regimes. For a fixed velocity resolution, a velocity channel will
remain \textit{thin} up to wavenumber
\begin{equation}
k \leq 2\pi \left[ \frac{1}{\sigma_L^2} \left( \delta v^2 + 2v_T^2
    \right) \right]^{-1/m} \ ,
\end{equation} 
where $\sigma_L$ is the velocity dispersion over the scale $L$,
equivalent to the size of the computational domain, and $v_T$ is the
thermal width. The thermal width smears velocity fluctuations on
smaller scales. 

\subsubsection{Projection Smoothing}
\label{subsec:projsmooth}
The effect of projecting a three-dimensional correlation function onto
a two-dimensional space has been studied by
\citet{{1951ZA.....30...17V}}, \citet{1958RvMP...30.1035M}, \citet{1987ApJ...317..686O}
and \citet{2004ApJ...604..196B}, among others. For the velocity field, the two-dimensional case corresponds to
the emission-line velocity centroid map. 

Previous authors have established
that for an isotropic, power-law, three-dimensional power spectrum, the
spectral index does not change on going from three dimensions to two
(projected) dimensions, i.e. $\kappa_{2D} = \kappa_{3D}$, where
$\kappa_{ND}$ is the power spectrum spectral index in $N$ dimensions
(defined as $p_N(k) \propto k^{-\kappa_{ND}}$, \citealp{{1998A&A...336..697S},{2003ApJ...593..831M},{2003ApJ...595..824B}}). This is because the line-of-sight contribution to
the line-of-sight velocity has no amplitude on the $k_z = 0$ plane,
where $z$ is the line-of-sight direction.

The relationship between the power spectrum spectral index and the
exponent of the second-order structure function for homogeneous
turbulence is $\kappa_{ND} = m_\mathrm{ND} + N$, where $m_\mathrm{ND}$
is the power-law index of the $N$-dimensional second-order structure
function. For projection from 3 to 2 dimensions, we therefore have
$m_\mathrm{2D} = m_\mathrm{3D} + 1$, and this is known as
\textit{projection smoothing}. 

In the case of incompressible, homogeneous (Kolmogorov) turbulence, we
have $\kappa_{3D} = 11/3$ and $m_\mathrm{3D} = 2/3$, hence the
relationship $\kappa_{2D} = \kappa_{3D}$ leads to $m_\mathrm{2D} =
5/3$. For the compressible turbulence \citep{1974Burgers} case, we
have $\kappa_{3D} = \kappa_{2D} = 4$ and $m_\mathrm{3D} = 1$, leading
to $m_\mathrm{2D} = 2$.  When the line-of sight depth is small compared
to the plane-of-the sky size of the emitting region, the distribution
of emittors is essentially two-dimensional (sheet-like) and we would
expect $m_\mathrm{2D} \simeq m_\mathrm{3D}$
\citep{{1987ApJ...317..686O},{2003ApJ...593..831M}}.

In order to compensate for the effects of density inhomogeneities,
\citet {2004ApJ...604..196B} introduced a correction term $\delta\kappa$
such that $\kappa_{2D} = \kappa_{3D} + \delta\kappa$, where
$\delta\kappa$ tends to $-1$ when density fluctuations are
important. This idea was applied by \citet{2011MNRAS.413..721L}  in an
effort to use 2D observational statistics to infer the 3D velocity
field of photoionized gas in giant extragalactic \hii{} regions.

\section[]{Results}
\label{sec:results}
\subsection{General Properties}
\label{subsec:genprop}
\begin{figure*}
\centering
\includegraphics[width=0.45\textwidth]{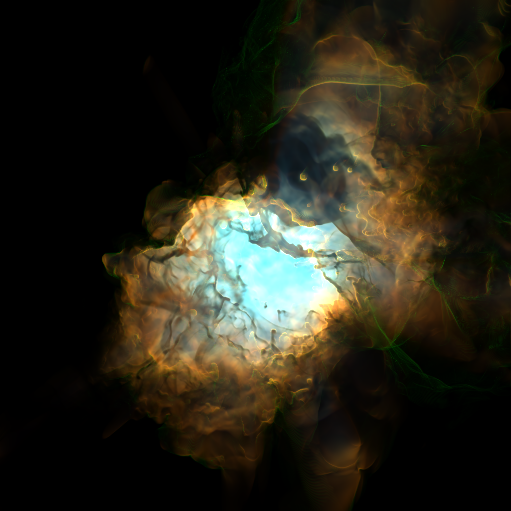}~\includegraphics[width=0.45\textwidth]{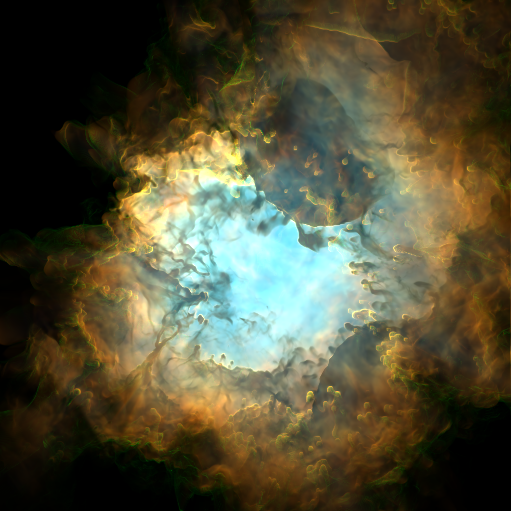}\\
\includegraphics[width=0.45\textwidth]{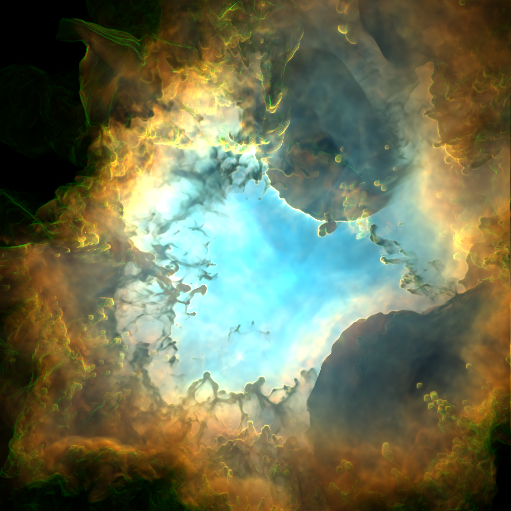}~\includegraphics[width=0.45\textwidth]{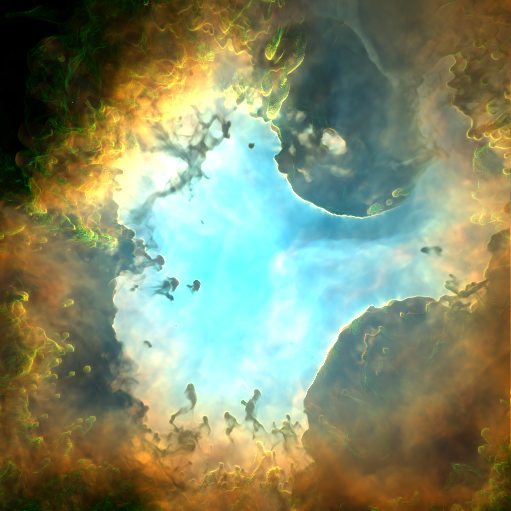}
\caption{Simulated projected emission-line images of an evolving \hii{} region in a
turbulent molecular cloud. The panels show the \hii{} region at 150,000,
200,000, 250,000 and 300,000~yrs, from top left to bottom right. The
physical size of each image is 4~pc on a side and the resolution of
the simulation is $512^3$ cells. The view is from the negative$-z$
direction of the computational cube. The colours represent the optical
emission lines \nii$\lambda 6584$ (red), H$\alpha$ (green), and
\oiii$\lambda 5007$ (blue), and for these images the effect of dust
absorption has been included in the radiative transfer during the
projection onto the plane of the sky. }
\label{fig:HIIimages}
\end{figure*}

We begin by considering the morphological appearance of the simulated
\hii{} regions. Figure~\ref{fig:HIIimages} shows the evolutionary
sequence from 150,000 to 300,000~years. Each image is a composition of
the three optical emission lines \nii$\lambda 6584$ (red), H$\alpha$
(green) and \oiii$\lambda 5007$ (blue), where we have employed the
classical \textit{Hubble Space Telescope} red-green-blue colour
scheme. Emission from the dense, neutral zones is negligible and for
the production of these images we
also include dust extinction in the radiative transfer for the
projection to the plane of the sky, with the assumption that the
dust-to-gas mass ratio is $1 \%$. 

The images show that the ionized gas distribution is not spherical and
around the corrugated edge of the \hii{} region there are many neutral
clumps and fingers of gas. This is a consequence of the \hii{} region
evolving in a clumpy neutral medium. The fingers and clumps at the
edge of the \hii{} region are the remnants of denser regions in the
initial turbulent cloud.  We can also see that the different ions are
important in different regions of the photoionized gas. The
\oiii$\lambda 5007$ (blue) emission is strongest closer to the
photoionizing source, while the \nii$\lambda 6584$ (red) emission is
most prominent around the edge of the \hii{} region and the H$\alpha$ is
distributed throughout the nebula. Although not shown in
Figure~\ref{fig:HIIimages}, we also calculate the \sii$\lambda 6716$
emission, which is important only around the edge of the \hii{} region in
the vicinity of the ionization front.  This is consistent with the
phenomenon of ionization stratification \citep{{2006agna.book.....O}}.

The neutral clumps and fingers around the edge of the \hii{} region are
the source of photoevaporated flows, which flow away from the
ionization front into the \hii{} region \citep{{2003RMxAC..15..175H}}. These flows
are mildly supersonic and can reach velocities of up to two or three
times the sound speed in the ionized gas. They shock against each
other in the interior of the \hii{} region. Neutral clumps inside the \hii{}
region are accelerated outwards by the rocket effect in the opposite
direction to the photoevaporated material flowing off their ionized
skins. In Figure~\ref{fig:HIIimages} these photoevaporated flows can
be discerned as shadowy regions ahead of the convex bright cusps that
mark the position of the ionization front. As the photoevaporated gas
flows away from the ionization front, the density drops in the
diverging flow and this is why the flows appear darker than the
surrounding photoionized gas.

\begin{figure}
\centering
\includegraphics[width=\linewidth]{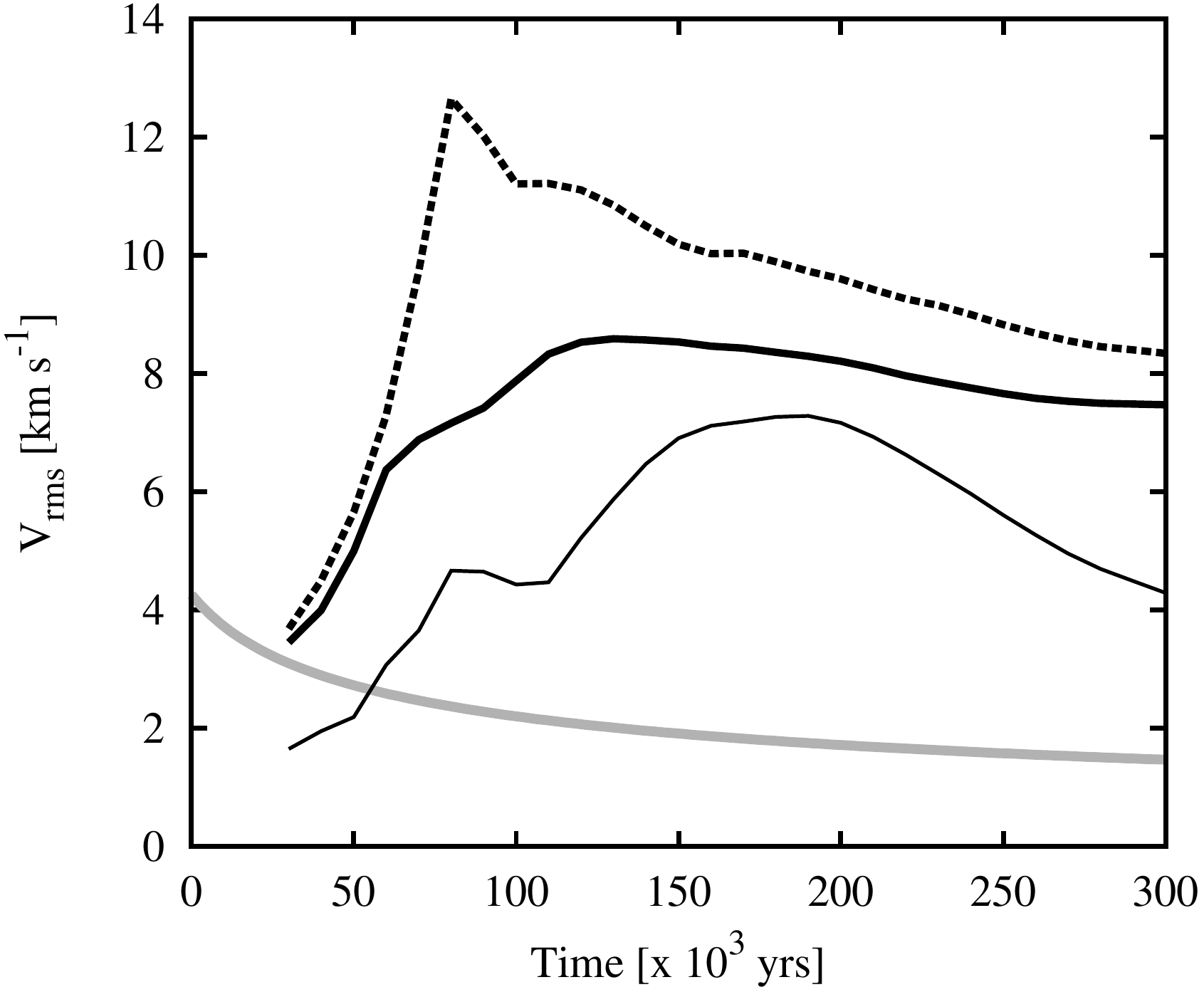}
\caption{Velocity dispersion as a function of time. The thick, black solid
  line shows the mass-weighted velocity dispersion of the ionized gas,
  the dotted line is the volume-weighted  velocity dispersion of the 
  ionized gas, while the thick grey line is the r.m.s.\ velocity of
  the analytical expansion \citep{1978Spitzer} in a uniform medium of
  density $n_0$. The thin, black continuous line is the 
  mass-weighted mean radial expansion velocity of the ionized gas.}
\label{fig:veldisp}
\end{figure}

These colliding photoevaporated flows are responsible for the velocity
dispersion of the ionized gas, which we show in
Figure~\ref{fig:veldisp}. This figure shows both the mass-weighted velocity
dispersion of the ionized gas and the
volume-weighted velocity dispersion. At early times ($t < 10^5$~yrs) the
volume-weighted velocity dispersion is higher than that of the
mass-weighted velocity dispersion. This is because the \hii{} region
breaks out of the dense clump where it is formed after about
50,000~yrs and a low-density, relatively high velocity champagne flow
results. After about 100,000~yrs, the main ionization front has just
about caught up and thereafter the two different velocity dispersions
show the same behaviour, remaining roughly constant with
$V_\mathrm{rms} \sim 8$~km~s$^{-1}$. In Figure~\ref{fig:veldisp} we also show the
r.m.s.\ velocity for the analytical  expansion \citep{1978Spitzer} in a uniform
medium, where the gas velocity behind the ionization front is one half
that of the ionization front and the internal velocities are linear
with radius. In this case, the velocity dispersion falls steadily as a
function of time, and after 100,000~yrs is less than 2~km~s$^{-1}$. The
differences between the velocity dispersions in the numerical and analytic
cases are mainly due to the interaction of the photoevaporated
flows produced in the simulations described above. In the analytic expansion, the velocities are radial
away from the central source. In the clumpy medium, the
photoevaporated flows lead to large non-radial velocities, which increase
the velocity dispersion in the photionized region. Finally,
Figure~\ref{fig:veldisp} also shows that the mass-weighted mean radial
velocity peaks just before 200,000~yrs at about 7~km~s$^{-1}$ and
thereafter falls off. This indicates that the global expansion will
not be an important influence on the dynamics, compared to the velocity
dispersion, after 200,000~yrs.
\begin{figure}
\centering
\includegraphics[width=\linewidth]{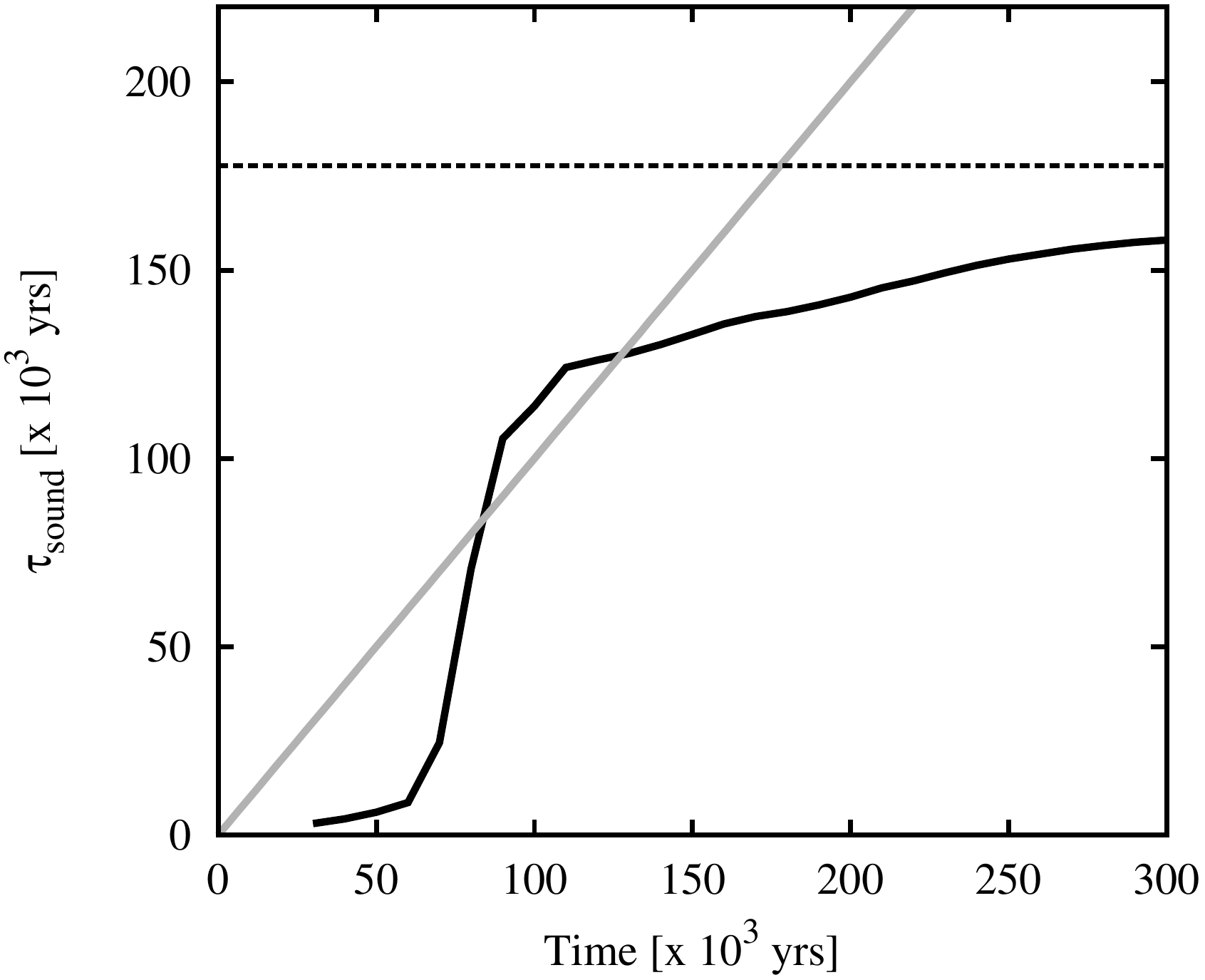}
\caption{Sound-crossing time as a function of evolution time. The
  sound-crossing time is $\langle R \rangle / c_\mathrm{i}$, where
  $\langle R \rangle$ is the radius of an equivalent sphere with
  volume equal to that of the \hii{} region and $c_\mathrm{i} =
  11$~km~s$^{-1}$ is the sound speed in the photoionized gas. The
  diagonal grey line has slope of unity. The horizontal dashed line is
  the sound-crossing time for a distance of 2~pc, which is the
  half-side length of the computational box.}
\label{fig:scross}
\end{figure}

In Figure~\ref{fig:scross} we compare the sound-crossing time of the
ionized volume with the evolution time of the simulation. During the
breakout of the \hii{} region from its initial dense clump (between
about 80,000 and 100,000~yrs), the expansion
is supersonic and the sound-crossing time exceeds the evolution
time. This coincides with the peak in the velocity dispersion (see
Fig.~\ref{fig:veldisp}). After this time, the expansion is subsonic and
the sound-crossing time becomes
shorter than the evolution time. A statistically steady state then
becomes possible for the turbulence within the ionized volume.

\subsection{Statistical Properties}
\label{subsec:statprop}
\subsubsection{Power Spectra}
\label{sssec:pspec}
\begin{figure*}
\centering
\includegraphics[width=\textwidth]{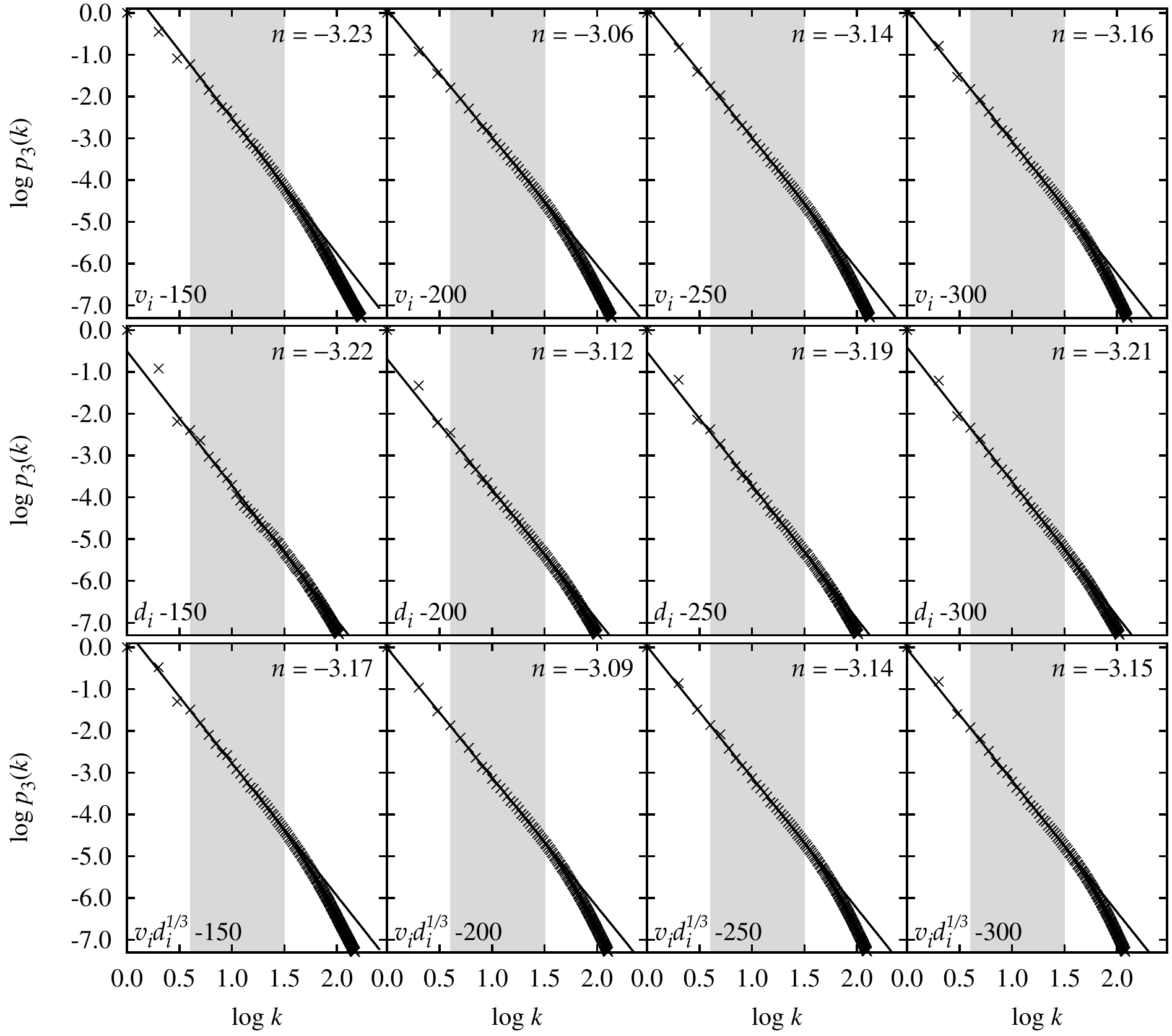}
\caption{Power spectra of the 3D simulation cube. From top to bottom: ionized gas velocity
  $v_i$, ionized gas density $d_i$, ionized gas velocity weighted by the cube root
  of the density $v_i d_i^{1/3}$.  From left to right: 150,000, 200,000, 250,000 and
  300,000~years. The points represent the calculated power spectra for
  the numerical simulation  and the solid line is the least-squares fit to
the data points between limits described in the
text, represented by the grey rectangle. The index $n$ of each
power-law fit is indicated in the corresponding panel.}
\label{fig:ps}
\end{figure*}
\begin{figure}
\centering
\includegraphics[width=\linewidth]{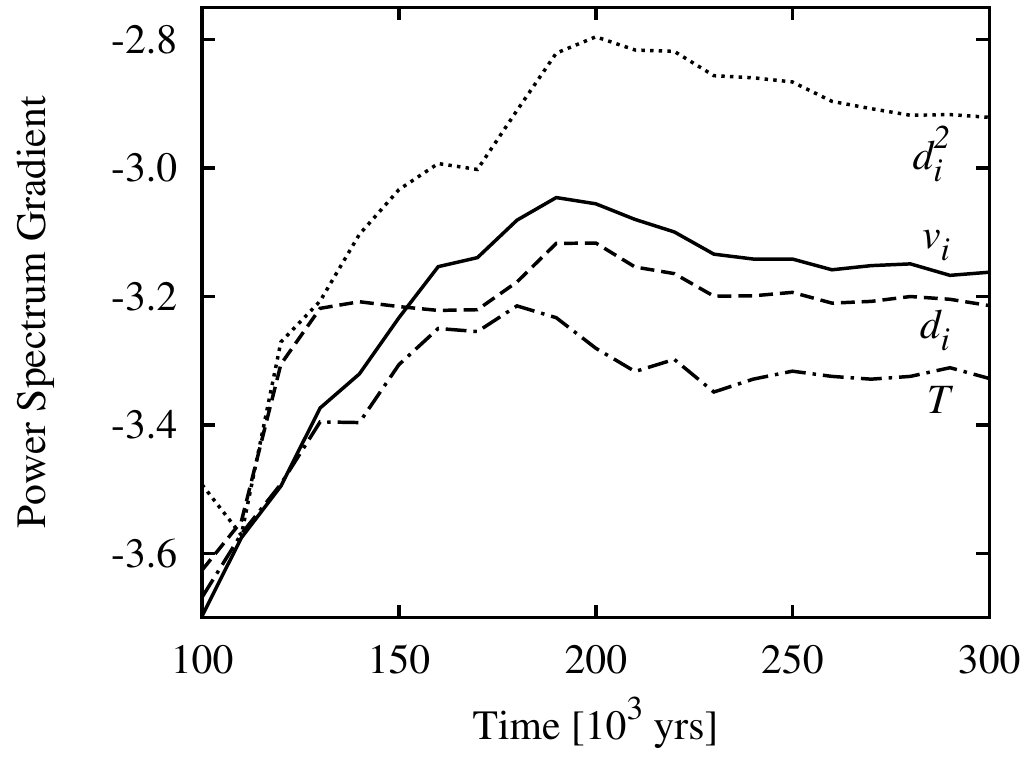}\\
\includegraphics[width=\linewidth]{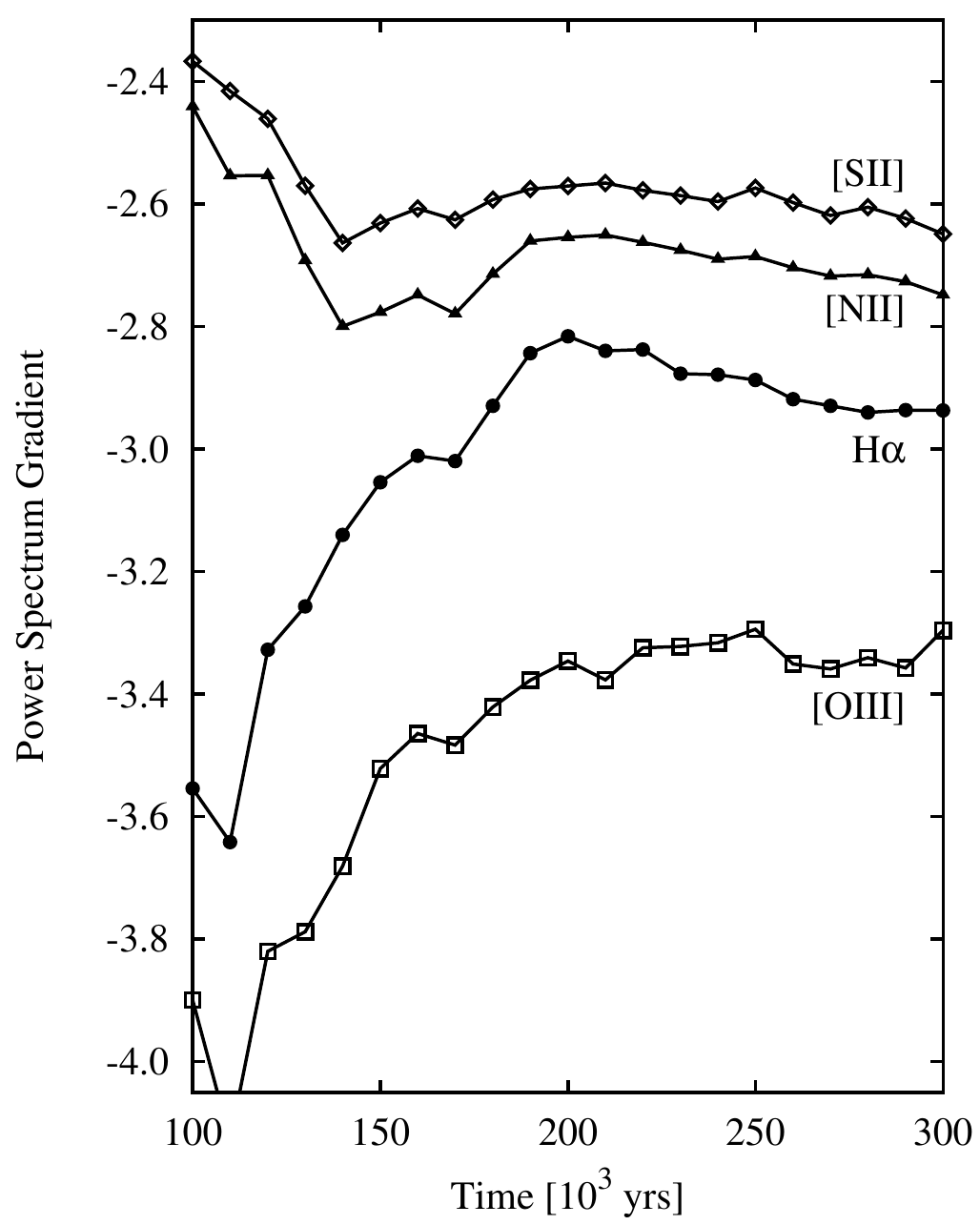}
\caption{Time variation of 3D power spectra power-law indices. Top panel: 3D physical
  quantities; solid line---ionized
  gas velocity $v_i$, dashed line---ionized density $d_i$, dotted line---square of ionized density $d_i^2$,
  dot-dashed line---temperature $T$. Bottom panel: emission-line
  volume emissivities; filled circles---$j(\mathrm{H}\alpha)$, open
  squares---$j(\mathrm{[OIII]})$, filled
  triangles---$j(\mathrm{[NII]})$, open diamonds---$j(\mathrm{[SII]})$.}
\label{fig:psevol}
\end{figure}

In Figure~\ref{fig:ps} we show the 3D power spectra of the ionized gas
velocity, the ionized density, and the ionized velocity weighted by
the cube root of the density \citep[see,
e.g.,][]{{2007ApJ...665..416K}}.  In the following sections we use a
dimensionless \(k\) that is normalised to the size of our
computational grid.  Thus \(k = 1\) corresponds to a physical scale of
\(4\)~pc.  The power spectra all exhibit a break in the power law at
about wavenumber $k \sim 32$, equivalent to a scale of 16
computational cells and consistent with the numerical dissipation
scale. We fit power laws between $k = 4$ (corresponding to a length
scale of one parsec) and $k =32$. For times later than 200,000~yrs,
the power-law fits for a given quantity are essentially constant (see Fig.~\ref{fig:psevol}). Note that here we use
$n$ to represent the 3D power spectrum spectral index,%
\footnote{This is to facilitate comparison with the theoretical
  literature, such as \citet{2000ApJ...537..720L}.}
which corresponds to the slope or gradient in log-log space, 
where $n = -\kappa_{3D}$ and $\kappa_{3D}$ is defined in
\S~\ref{subsec:projsmooth}. At earlier times, the power laws are
steeper and evolving. 
The settling down of the power spectra power-law indices
correlates with the time at which the global expansion ceases to be
important for the internal dynamics (see Fig.~\ref{fig:veldisp}).
This occurs after about 200,000~yrs (as evidenced in Fig.~\ref{fig:psevol}), which
corresponds to the evolution time being equal to approximately 1.5~times the
sound-crossing time (see Fig.~\ref{fig:scross}).

\subsubsection{Second-order Structure Functions}
\label{sssec:s2func}
\begin{figure*}
  \centering
  \includegraphics[width=0.8\textwidth]{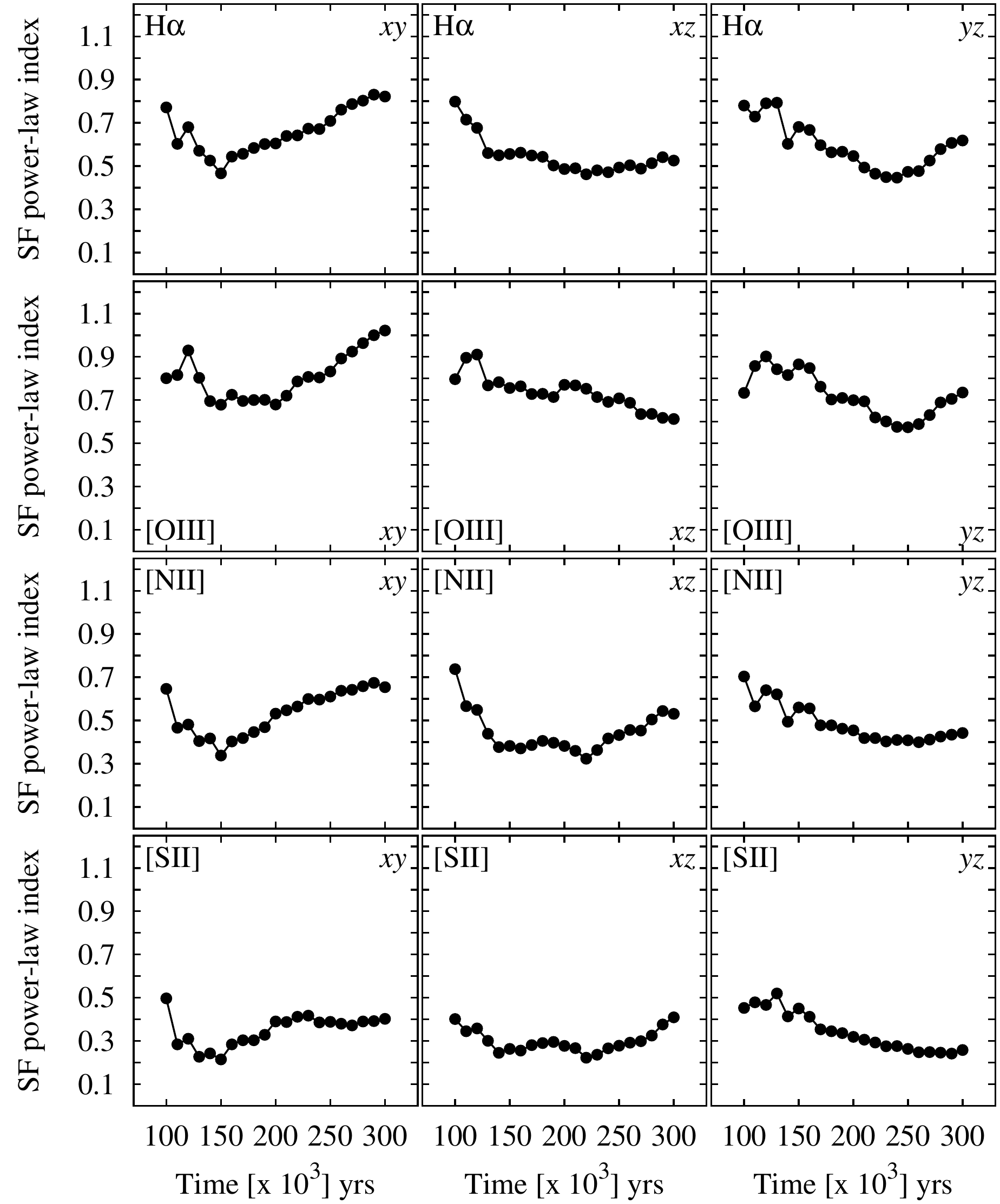}
  \caption{Evolution of second-order structure function power-law
    index, \mSF, as a function of time. From top to bottom: H$\alpha$,
    \oiii{} $\lambda 5007$, \nii{} $\lambda 6584$, \sii{} $\lambda
    6716$. From left to right: line of sight along the $z$, $y$ and
    $x$ axes, respectively.}
  \label{fig:sftrends}
\end{figure*}

\begin{figure*}
\centering
\includegraphics[width=0.6\linewidth]{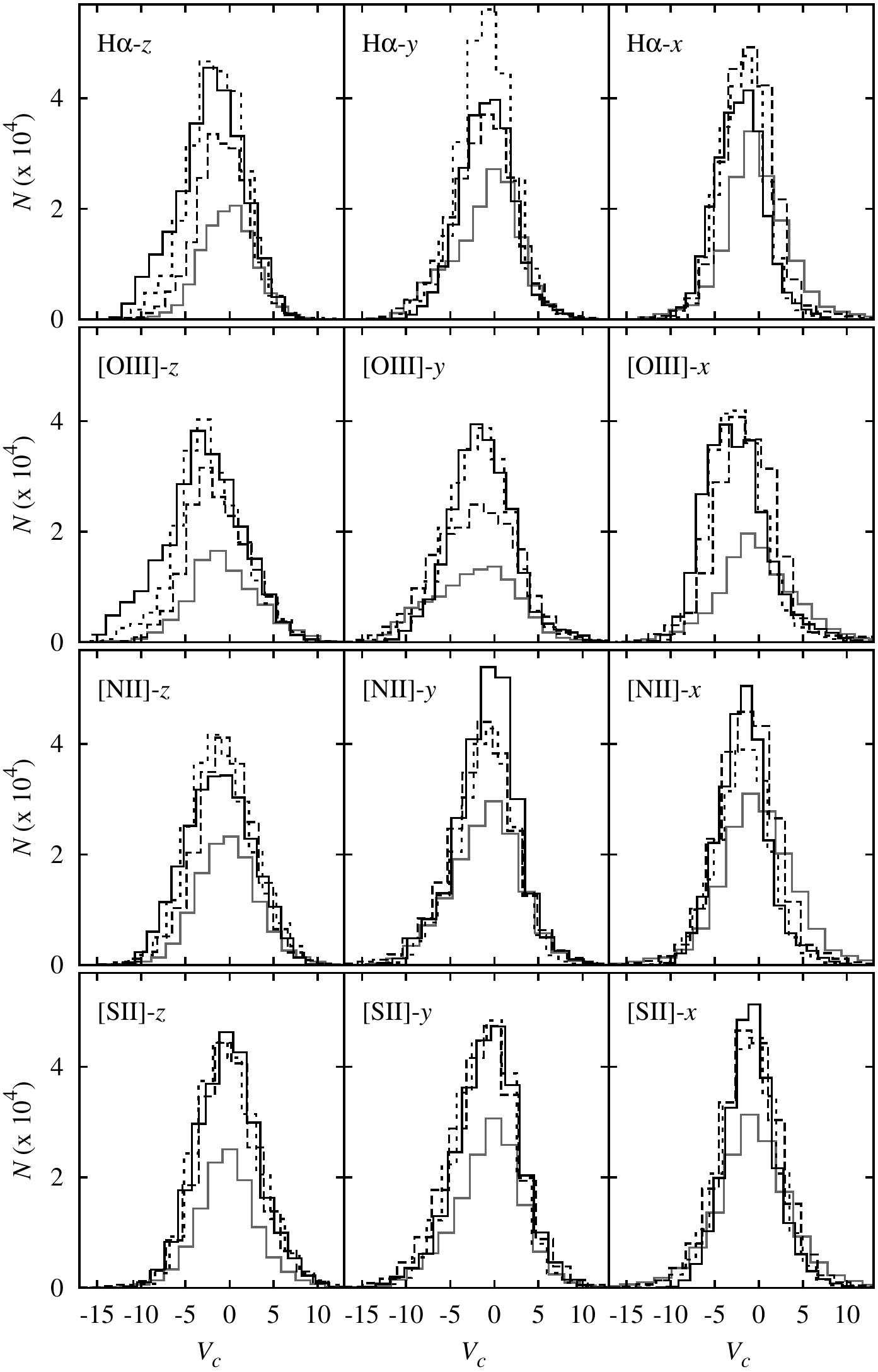}
\caption{Histograms of velocity centroid values for each emission line
  along different lines of sight. From top to bottom: H$\alpha$,
  \oiii$\lambda 5007$, \nii$\lambda 6584$, \sii$\lambda 6716$. From
  left to right: line of sight along the $z$, $y$ and $x$ axes,
  respectively. The different line types refer to different times:
  thick, grey line---150,000~yrs, dashed line---200,000~yrs,
  short-dashed line---250,000~yrs, continuous black
  line---300,000~yrs.}
\label{fig:histogram}
\end{figure*}

We use the procedure described in Section~\ref{subsubsec:centroid} to
calculate velocity centroid maps for the H$\alpha$, \oiii$\lambda
5007$, \nii$\lambda 6584$ and also \sii$\lambda 6716$ emission lines
and then calculate the corresponding second-order structure functions
according to Equation~\ref{eq:strucfunc}. Results for representative
evolutionary times are shown in Figures~\ref{fig:sfunc} to
\ref{fig:sfuncyz} of Appendix~\ref{app:sf}, where fits to the
power-law index (\mSF) of the structure function, which corresponds to
the slope or gradient in log-log space, are carried out for the
inertial range of scales. A description of the procedure for
identifying the inertial range is given in Appendix~\ref{app:sf} and
illustrated by the accompanying Figure~\ref{fig:sfauto}.

In Figure~\ref{fig:sftrends} we show the evolution of \mSF{} with time
for the different lines and for the three principal viewing directions
of the simulation cube.  For the line of sight along the $z$-axis
(first column of Fig.~\ref{fig:sftrends}), one sees for all lines a
consistent steepening of the structure function graph with time
(increase in \mSF{}).  But for other viewing directions no such trend
is apparent: both rising and falling behavior of \mSF{} is seen, with
little consistency between different lines.

In order to understand why one particular viewing direction is
different, we produced histograms of the emission-line velocity
centroid values binned into narrow $<2$~km~s$^{-1}$ bins for the three
different lines of sight at the four different times. The histograms
are presented in Figure~\ref{fig:histogram}, from which we see that
for the $z$-axis line of sight, the values of $V_c$ are not
distributed symmetrically about the mean value and, in fact, for the
H$\alpha$ and \oiii$\lambda$5007 emission lines, a ``wing'' develops
for negative values of $V_c$ that extends to more negative values as
time progresses. This tendency is not seen for the $y$- and $x$- axis
lines of sight. We attribute this wing to a ``champagne'' flow towards
the observer along the $z$-axis. This flow would be perpendicular to
the line of sight for observations along the other axes.

\subsubsection{Velocity Channel Analysis}
\label{sssec:vca}
\begin{figure*}
  \centering
  \includegraphics[width=\linewidth]{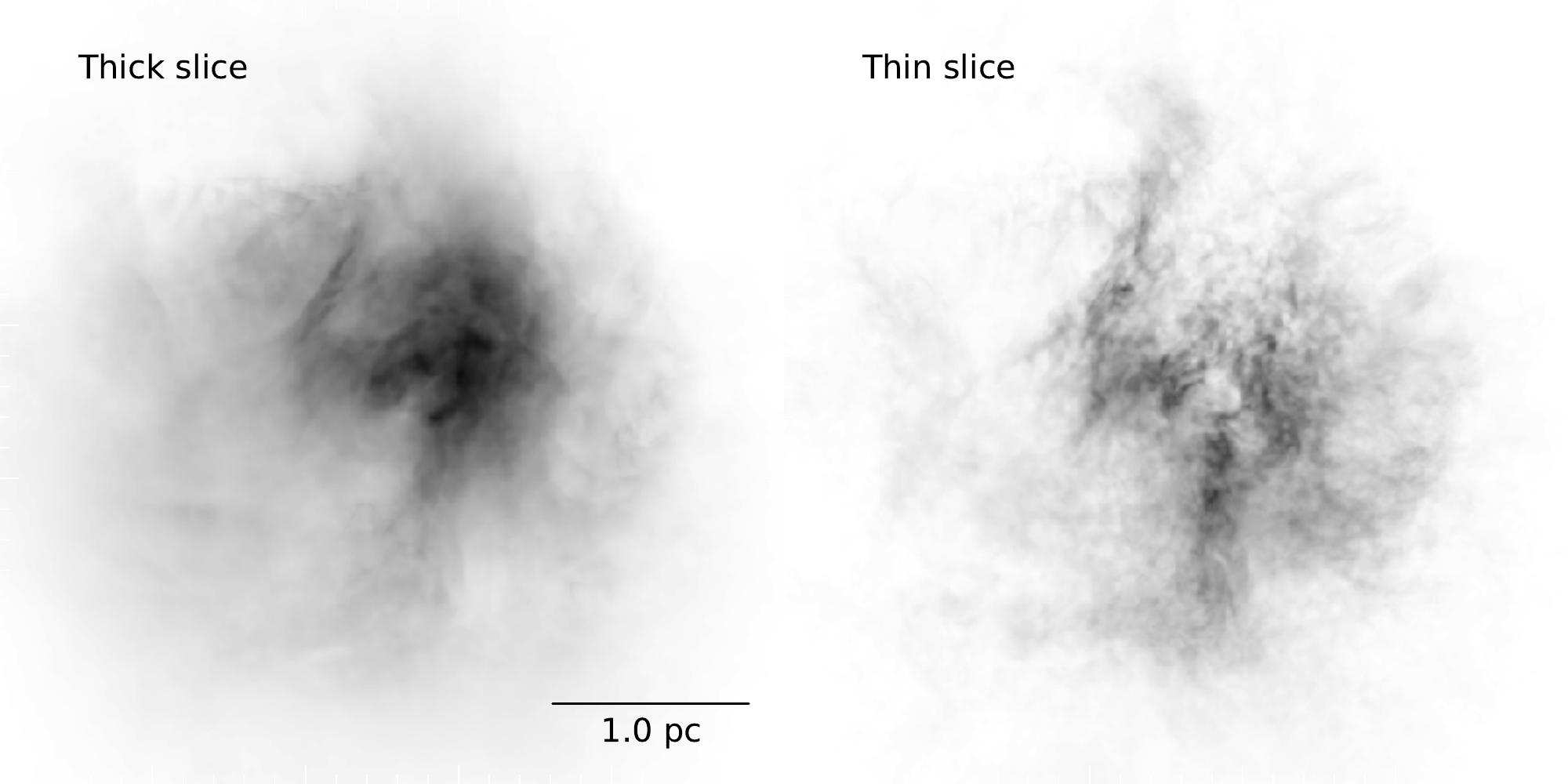}
  \caption{Surface brightness maps in thick (left) versus thin (right)
    velocity slices for the \oiii{} line from our simulation at an age
    of 300,000~years.  The thick slice covers the full velocity range
    of the emission line, while the thin slice has a width of
    5~km~s$^{-1}$, which is smaller than the turbulent velocity
    fluctuations, but slightly larger than the thermal broadening for
    this line.  It is apparent that the thin slice shows considerably
    greater small-scale structure than the thick slice, which is
    reflected in its shallower power spectrum.  The brightness
    structure in the thick slice is due entirely to the emissivity
    fluctuations within the \hii{} region, whereas the additional
    structure in the thin slice is caused by velocity fluctuations.  }
  \label{fig:o3-thick-thin}
\end{figure*}

\begin{figure*}
\centering
\includegraphics[width=0.8\textwidth]{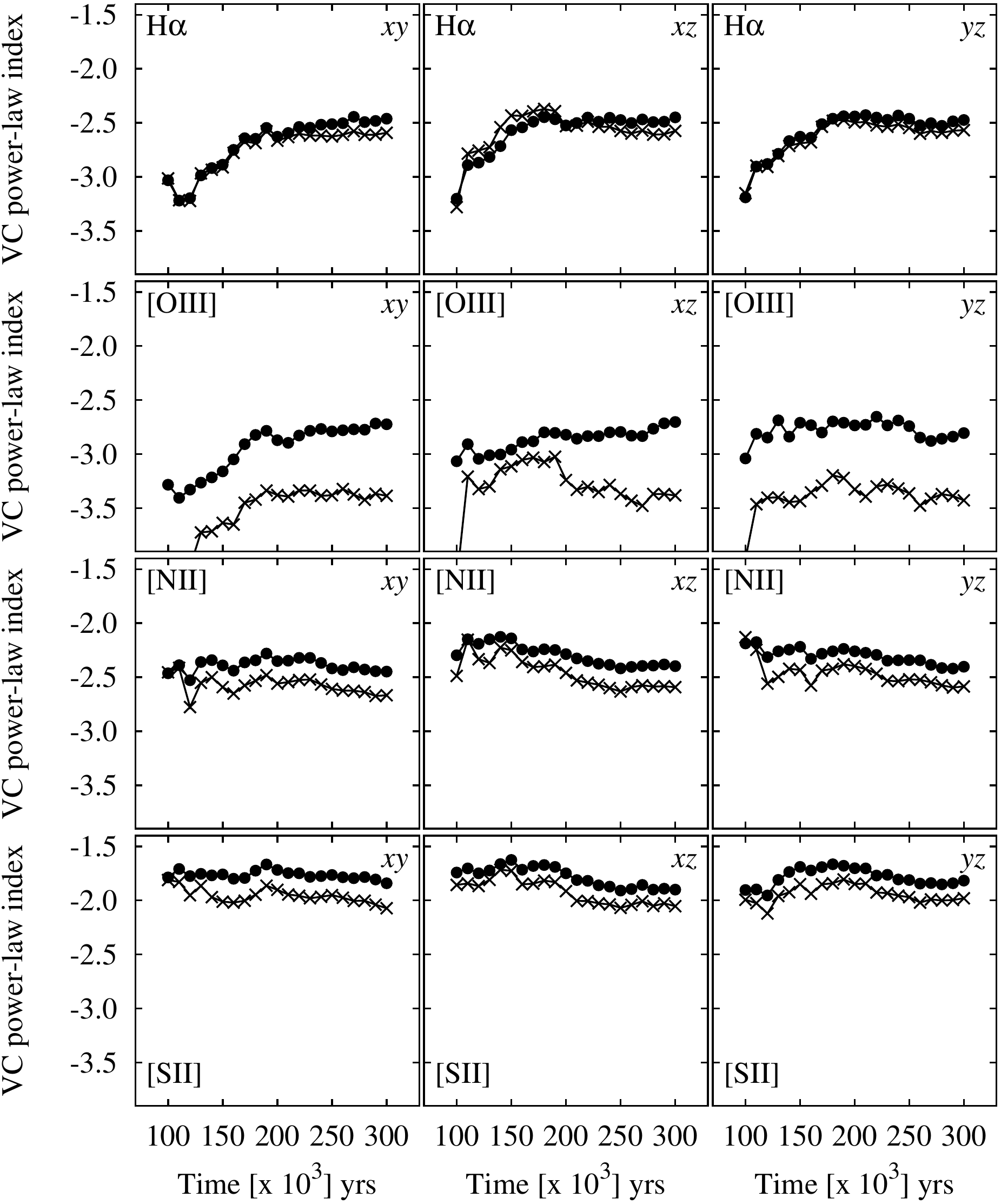}
\caption{ Evolution of velocity channel power-law index as a function
  of time for thick channels (\gammaVCAvthick; crosses) and thin
  channels (\gammaVCAthin; filled circles).  From top to bottom:
  H$\alpha$, \oiii{} $\lambda 5007$, \nii{} $\lambda 6584$, \sii{}
  $\lambda 6716$. From left to right: line of sight along the $z$, $y$
  and $x$ axes, respectively.  Thermal broadening was included in all
  cases.  }
\label{fig:vcatrends}
\end{figure*}

The velocity channel analysis consists of calculating the
two-dimensional power spectrum of the brightness distribution
in isovelocity channels of varying thickness.  
We consider two cases: thick slices,
which are wide enough (\(\sim 100~\mathrm{km\ s^{-1}}\))
to include all the emission in the line,
and thin slices, with width \(5~\mathrm{km\ s^{-1}}\). 
Because the velocity spectrum in our simulations is rather shallow (see above),
the line-of-sight turbulent velocity dispersion \(\delta v\)
exceeds the width of these thin slices
over the full range of length scales that we can usefully study,
from \(0.1\)~pc (\(\delta v \approx 8~\mathrm{km\ s^{-1}}\))
to \(1\)~pc (\(\delta v \approx 10~\mathrm{km\
  s^{-1}}\)). Figure~\ref{fig:o3-thick-thin} shows typical examples of
the \oiii{} brightness in thick and thin slices.

To use thinner slices would not be useful for a variety of reasons.
First, \(5~\mathrm{km\ s^{-1}}\) corresponds to the highest resolution 
that can be achieved with optical spectrographs
that are optimised for studying extended sources,
such as Keck HIRES or VLT UVES. 
Second, thinner slices are increasingly subject to ``shot noise'' 
due to the finite resolution of the numerical simulations,
which produces spurious small-scale power, as discussed by 
\citet {2003MNRAS.342..325E} and \citet {2003ApJ...593..831M}.
Third, thermal broadening would smoothe out any structure on 
scales \(< 5~\mathrm{km\ s^{-1}}\) for all but the heaviest ions.

Figure~\ref{fig:vcatrends} shows the evolution with time of the VCA
power-law indices from thin (\gammaVCAthin{}) and thick (\gammaVCAvthick{}) channels (shown by filled circle and cross
symbols, respectively) for different ions and for different viewing
directions.  The individual VCA power spectra from which these
power-law indices, which correspond to the slope or gradient in log-log space,
were extracted are presented in Appendix~\ref{app:vca}.  It can be
seen that both \gammaVCAthin{} and \gammaVCAvthick{} are remarkably
stable with time during the latter part of the evolution (\(t >
200,000\)~years).  Although thermal broadening means that there is no
clear distinction between \gammaVCAthin{} and \gammaVCAvthick{} for
the H\(\alpha\) line, the two values are clearly distinguished for the
heavier ions, with the thin slices showing a significantly shallower
slope, especially for \oiii{}.  The implications for diagnosing
turbulence statistics are discussed in \S~\ref{sssec:vca2}.

\section{Discussion}
\label{sec:discuss}

\subsection{Characterization of the turbulence from optical emission
  lines}
\label{subsec:charac}
At times later than 150,000~years, our second-order structure function
results for the H$\alpha$ and \oiii$\lambda$5007
emission-line velocity centroids strongly suggest the presence of
turbulence with an inertial range between 1~pc and the numerical
dissipation scale of about 8 cells (equivalent to 0.0625~pc). The
\nii$\lambda$6584 and
\sii$\lambda$6716 structure function results also suggest turbulence, with a smaller upper limit to the inertial range,
which is consistent with these ions being confined to relatively thin
layers near the ionization front. However, it is difficult
to characterize this turbulence since the slope of the structure
function for a given emission line varies with time in an
unpredictable manner and also depends on the line of sight. Although
the results for the $z$-axis line of sight suggest an increase in
slope with time, an examination of the distributions of the velocity centroids
(see Fig.~\ref{fig:histogram}) shows that this is due to a champagne-type
flow in that direction, and other lines of sight do not show a
definite trend with time.

Different emission lines originate in different volumes of ionized
gas, and this is reflected in the different slopes for the structure
functions from different emitters. The H$\alpha$ line is produced
throughout the volume and is brightest close to the ionization front
around the bright edges of the photoionized gas. There is therefore a
wide range of densities associated with the H$\alpha$-emitting
gas. The emissivity of the H$\alpha$ recombination line depends on the
square of the density and only weakly on the temperature in the
photoionized gas \citep[see e.g.,][]{2006agna.book.....O}. Indeed, the
gradients of the 3D power spectra of the H$\alpha$ emissivity and
the square of the density are essentially the same. 
On the other hand, the emissivity of the \oiii$\lambda$5007
collisional line depends more strongly on temperature. This line
originates in the interior of the \hii{} region, where the density is
more uniform but weak shocks due to the collision of photoevaporated
flows cause fluctuations in the temperature.
A more uniform density distribution
corresponds to a steeper density power spectrum and, indeed, the
structure functions, 2D velocity channel power spectra, and 3D power
spectrum of the \oiii$\lambda$5007 line are all steeper than those
for the H$\alpha$ recombination line.

The stellar parameters for the simulations presented in this work
correspond to a relatively hot (37,500~K) O7 star. For these
parameters, the \nii$\lambda$6584 and \sii$\lambda$6716
collisionally ionized lines come from regions close to the ionization
front, where the density variations are strong, and this is reflected
in the less steep structure function and 2D velocity channel power
spectra gradients. In particular, the \sii{} emission will come from
very close to the ionization front where the acceleration of the
ionized gas is strongest, and as a result the structure function and
2D velocity channel power-spectra gradients are shallowest for this
line. Other stellar parameters, e.g., a cooler B0 star or a much
hotter white dwarf, would produce photoionized regions with 
different ionization stratifications.


\subsubsection{Intrinsic Power Spectra Of Physical Quantities}
\label{sssec:ips}
Figure~\ref{fig:ps} shows that
the power spectra of physical quantities are very well approximated
by power laws over the range from $k = 4$ to 32
(scales of 1~pc to 0.125~pc).
In particular, the ionized gas velocity shows
a power-law slope of \(n = -3.2 \pm 0.1\)
once the turbulence is fully developed. 
This is significantly shallower than the Kolmogorov (\(n = -3.667\))
or Burgers (\(n = -4\)) value,
indicating more velocity structure at small scales than would be seen
in a simple turbulent cascade of energy injected at the largest scale.
As a consequence,
the turbulent velocity dispersion is relatively insensitive to scale,
varying as \(\sigma \sim l^{0.5 (-3 - n)} \sim l^{0.1}\). 


One possible reason for the shallow velocity power spectrum may be
that energy is injected over a variety of scales, corresponding to the
different sizes of clumps and filaments responsible for the photoevaporated
flows in the simulated \hii{} regions. Moreover, the
energy injection will vary with time due to the global expansion of
the \hii{} region, which moves the sources of the photoevaporated
flows generally outwards, and the destruction of the clumps and
filaments as they are eroded by the ionizing radiation.

The density power spectrum has a very similar slope to that of the velocity: \(n = -3.2 \pm 0.1\),
but of greater relevance
are the slopes of the emissivities of the different emission lines,
which are \(n = -3.4 \pm 0.1\) for \oiii{}, 
\(n = -2.9 \pm 0.1\) for \ha,
\(n = -2.7 \pm 0.1\) for \nii, and
\(n = -2.6 \pm 0.1\) for \sii.
These span the critical value of \(n = -3\)
that divides ``steep'' from ``shallow'' power spectra.
\oiii{} has a steep slope,
indicating that large-scale fluctuations dominate,
while \nii{} and \sii{} have shallow slopes,
indicating that small-scale fluctuations dominate. 
The \ha{} slope is very close to the critical value,
indicating roughly equal contributions from fluctuations on all size-scales.

It is interesting to study the question of whether
the known power-law indices of the velocity and emissivity power spectra in our simulations
can in practice be recovered from observational diagnostics.  
If this is not the case for a given diagnostic,
then it would call into question its utility for studying real \hii{} regions.
In particular, we will concentrate on two commonly used diagnostics:
the second-order structure function of the line velocity centroids,
and the power spectra of the surface brightness in isovelocity channel maps
(Velocity Channel Analysis). 

\subsubsection{Structure Function}
\label{sssec:strfunc}
The structure function of the velocity centroids is an observationally
attractive diagnostic because it is relatively immune to the effects
of thermal broadening and poor spectral resolution, so long as
sufficiently high signal-to-noise spectra are used.  However, it has
the disadvantage that relating the observed slope to the 3-dimensional
velocity statistics depends on the geometry of the emitting region,
see \S~\ref{subsec:projsmooth}.  For transverse separations larger
than the characteristic line-of-sight depth of the emitting gas, the
two-dimensional gradient should be equal to the three-dimensional one:
\[
m_{\mathrm{2D}} = m_{\mathrm{3D}} = -3 - n,
\]
whereas at smaller separations than this, projection smoothing, as
described above, means that the two-dimensional gradient is steeper:
\[
m_{\mathrm{2D}} = 1 + m_{\mathrm{3D}} = -2 - n.
\]
Based on our simulation's velocity power spectrum index at late times
of \(n \approx -3.2\) (see Figs.~\ref{fig:ps} and \ref{fig:psevol}),
the structure function slope should be \(m_{\mathrm{2D}} = 0.2\) in
the large-scale limit and \(m_{\mathrm{2D}} = 1.2\) in the small-scale
limit.

In fact, all of the measured slopes lie between these two limits,
with a systematically increasing value from low to high-ionization lines:
\(m_{\mathrm{2D}}(\sii) = 0.33 \pm 0.02\), 
\(m_{\mathrm{2D}}(\nii) = 0.49 \pm 0.03\), 
\(m_{\mathrm{2D}}(\ha) = 0.59 \pm 0.04\), 
\(m_{\mathrm{2D}}(\oiii) = 0.74 \pm 0.04\), where the averages were
performed for \(t > 200,000\)~years. 
This is qualitatively consistent with expectations because the
emission from lower-ionization lines is confined to thin layers near
the ionization front, whereas higher ionization emission is more
distributed over the volume and therefore subject to greater
projection smoothing.

If the line-of-sight depth were constant over the face of the \hii{}
region, then the structure function would show a break at a scale
equal to that depth, but in reality the depth varies from point to
point, so there will not be a sharp break.  Instead, the structure
function is expected to show negative curvature, with the gradient
gradually decreasing as one passes from smaller to larger scales.  A
small such effect is seen in the structure functions derived from our
simulations (Fig.~\ref{fig:sfunc} to \ref{fig:sfuncyz}): the fit to a
power law is generally not so good as in the case of the power
spectra, with negative residuals at both ends of the fitted range,
indicative of a negative curvature.  That the observed effect is so
small is probably due to the fact that the distribution of
line-of-sight depths strongly overlaps with the limited dynamic range
in separations available from our simulations, bounded at small scales
by numerical dissipation, and at large scales by the size of the
ionized region.

It is disappointing that none of the measured slopes reach either of
the limiting cases discussed above.  All that can be deduced from the
structure function is that \(1 + m_{\mathrm{3D}} >
m_{\mathrm{2D}}(\oiii)\) and \(m_{\mathrm{3D}} <
m_{\mathrm{2D}}(\sii)\), which implies $n = -2.74$ to $-3.33$.
Although this is a rather wide range of allowed velocity power
spectrum slopes, it does serve to clearly rule out the Kolmogorov value of \(n
= -3.667\).   Furthermore, the ``true'' value of \(n = -3.12 \pm 0.03
\) lies close to the middle of the allowed range.  

A further proviso to the use of the structure function is that
systematic anisotropic flows can affect the measured slopes when the
viewing angle is along the direction of the flow.  Such an effect is
seen at later times for our simulation when viewed along the
\(z\)-axis (Fig.~\ref{fig:sfunc}).  In this case, the structure
function tends to steepen at the large-scale end of our fitting range,
producing a positive curvature, which is opposite to the more typical
case of negative curvature discussed above.  Such cases may also be
identified by the presence of a significant skew in the PDF of the
line-of-sight velocity (see Fig.~\ref{fig:histogram}).

\begin{figure*}
  \centering
  \includegraphics[width=\linewidth]{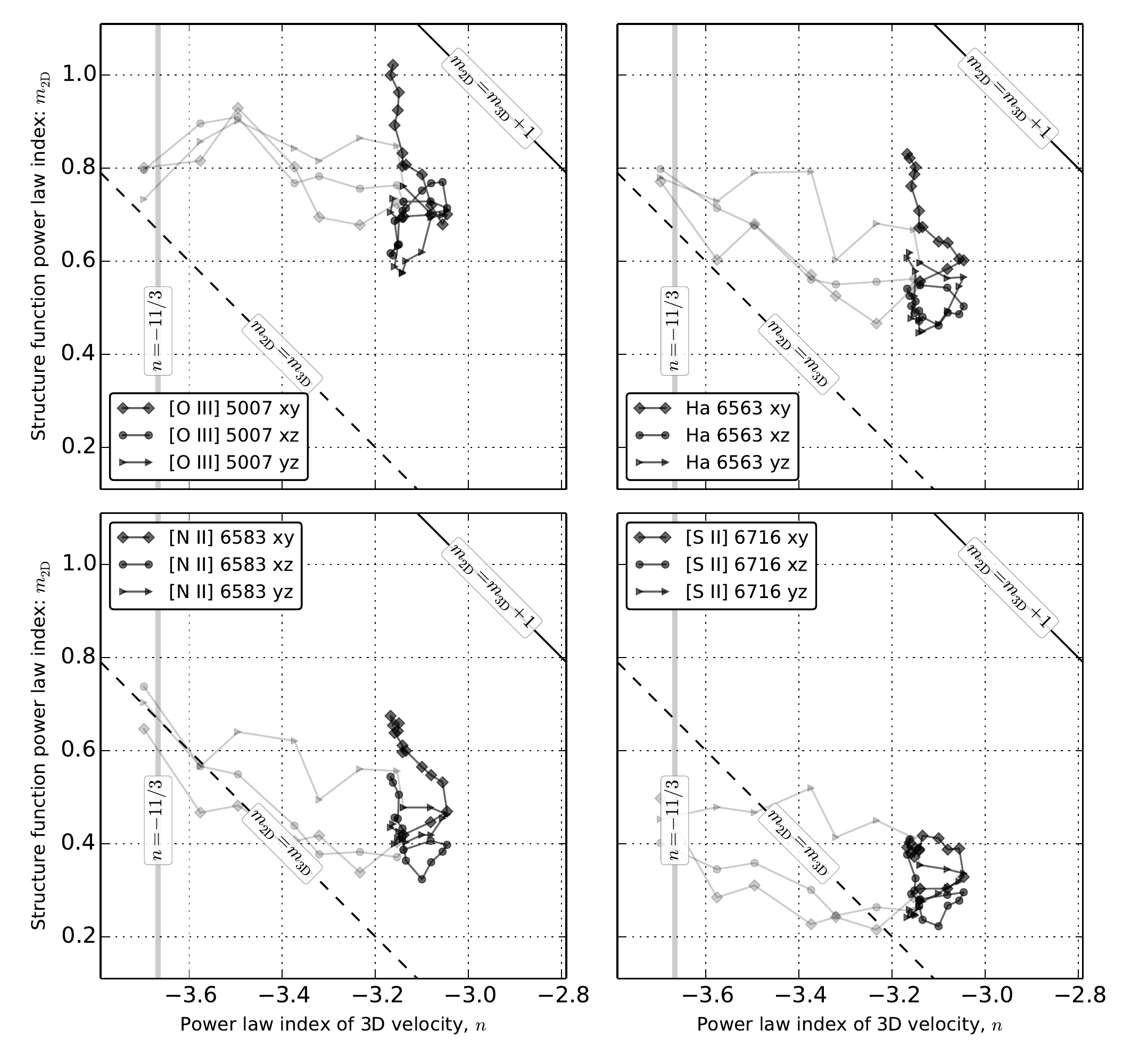}
  \caption{Structure function slope versus velocity power law slope.  Each panel shows a different emission line; clockwise from upper left: \oiii{}, \ha{}, \sii{}, \nii{}.  Structure function slopes are shown for the 3 principal viewing directions, distinguished by different symbol types (see key).  Dim gray lines and symbols show evolutionary times \(< 200,000\)~years, while black lines and symbols show times \(> 200,000\) years.} 
  \label{fig:sf-vs-n}
\end{figure*}

Figure~\ref{fig:sf-vs-n} illustrates these points by graphing the
correlation between the structure function slope \mSF{} and the slope
\(n\) of the underlying 3D velocity fluctuations.  The theoretical
relation is shown by black diagonal lines, both with (continuous line) and
without (dashed line) projection smoothing.  It is apparent that a
large part of the variation in \mSF{} is not driven by changes in
\(n\).  Indeed, \mSF{} shows a larger or equal variation in the latter
stages of evolution, when \(n\) is approximately constant, than it
does in the earlier stages, when \(n\) is varying.

Note that the additional complication identified by \citet
{2004ApJ...604..196B}, whereby correlations between density and
velocity fluctuations affect the translation between
\(m_{\mathrm{2D}}\) and \(n\), is likely of minor importance in our
case.  \citet {2007MNRAS.381.1733E} show that this is most important
for high Mach number turbulence, where \(\delta\rho/\langle \rho
\rangle > 1\), whereas the transonic turbulence inside our simulated
\hii{} regions produces more modest density contrasts.

\subsubsection{Velocity Channel Analysis}
\label{sssec:vca2}

\begin{figure*}
  \centering
  \includegraphics[width=\linewidth]{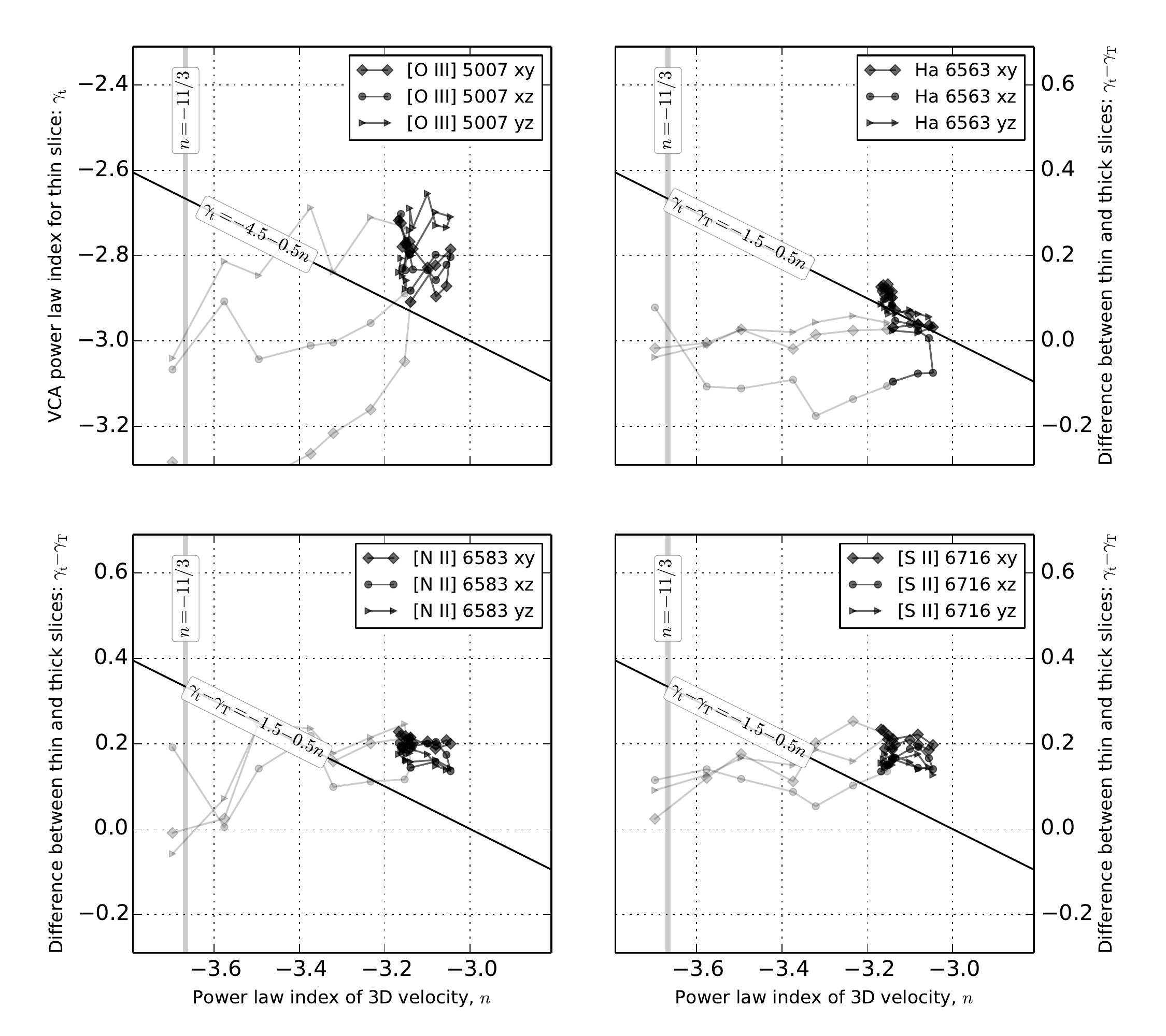}
  \caption{As Fig.~\ref{fig:sf-vs-n}, but showing VCA power spectrum
    slopes for thin slices, \gammaVCAthin{} or \(\gammaVCAthin -
    \gammaVCAvthick\), versus slope, \(n\), of the intrinsic velocity
    power spectrum for each emission line.  The \oiii{} line (upper
    left panel) has a sufficiently steep emissivity power spectrum
    that it is the absolute value of \gammaVCAthin{} that is predicted
    to be related to \(n\), as shown by the diagonal solid line.  The
    other emission lines have a shallower emissivity power spectrum,
    such that it is the relative slope between the thin and thick
    slices, \(\gammaVCAthin - \gammaVCAvthick\) that is predicted to
    depend on \(n\). }
  \label{fig:vca-thin-vs-n}
\end{figure*}

Figure~\ref{fig:vca-thin-vs-n} shows the correlations between the
slope of the velocity fluctuation power spectrum and the VCA slopes
found in \S~\ref{sssec:vca} above (see Fig.~\ref{fig:vcatrends}).  The
theoretical procedure \citep{2000ApJ...537..720L} for deriving one
from the other is slightly different, depending on whether the power
spectrum of the emissivity fluctuations is ``steep'' or ``shallow''
(see \S~\ref{sec:stats-vca} above).  In the steep case, which
applies to \oiii{} in our simulation, the slope of the average power
spectrum of the brightness maps in the thin isovelocity channels is
given by \(\gammaVCAthin{} = -3 + \frac12 m_{\mathrm{3D}}\),
where \(m_{\mathrm{3D}} = -3 - n = 0.2 \pm 0.1\) for our simulation.
The derived value from the \oiii{} thin channel maps for \(t >
200,000\) is \(\gammaVCAthin{} = -2.80 \pm 0.07 \), which
compares well with the value \(-2.9 \pm 0.05\) that is implied by the
simulation's value of \(n\).

In the shallow case, it is the difference in slope
between the thin and thick slices
that is predicted to depend on the velocity fluctuations:
\(\gammaVCAthin{} - \gammaVCAvthick{} = \frac12 m_{\mathrm{3D}}\). 
The derived values are 
\(\gammaVCAthin{} - \gammaVCAvthick{} = 0.07 \pm 0.05\), 
\(0.19 \pm 0.02\), and \(0.17 \pm 0.02\)
for \ha, \nii, and \sii, respectively. 
These also compare tolerably well with the value of \(0.1 \pm 0.05\)
that is implied by the simulation's value of \(n\).  

Note, however, that the large Doppler width of the \ha{} line means
that the thin velocity slices are not useful in this case, since the
thick and thin slices have identical slopes. The fact that this agrees
with the theoretical prediction is merely a coincidence, due to our
velocity spectrum having a slope that is close to \(-3\).  For the
lines from heavier ions, \oiii{}, \nii{} and \sii{}, the difference
between the thin and thick velocity slices is not erased by thermal
broadening, but in these three cases there is a consistent difference
of \(\approx 0.1\) between the measured VCA slope and the
theoretically expected one.  The origin of this difference is unclear,
but it is small enough that it is not a significant impediment to the
application of the VCA method.

\begin{figure*}
  \centering
  \includegraphics[width=\linewidth]{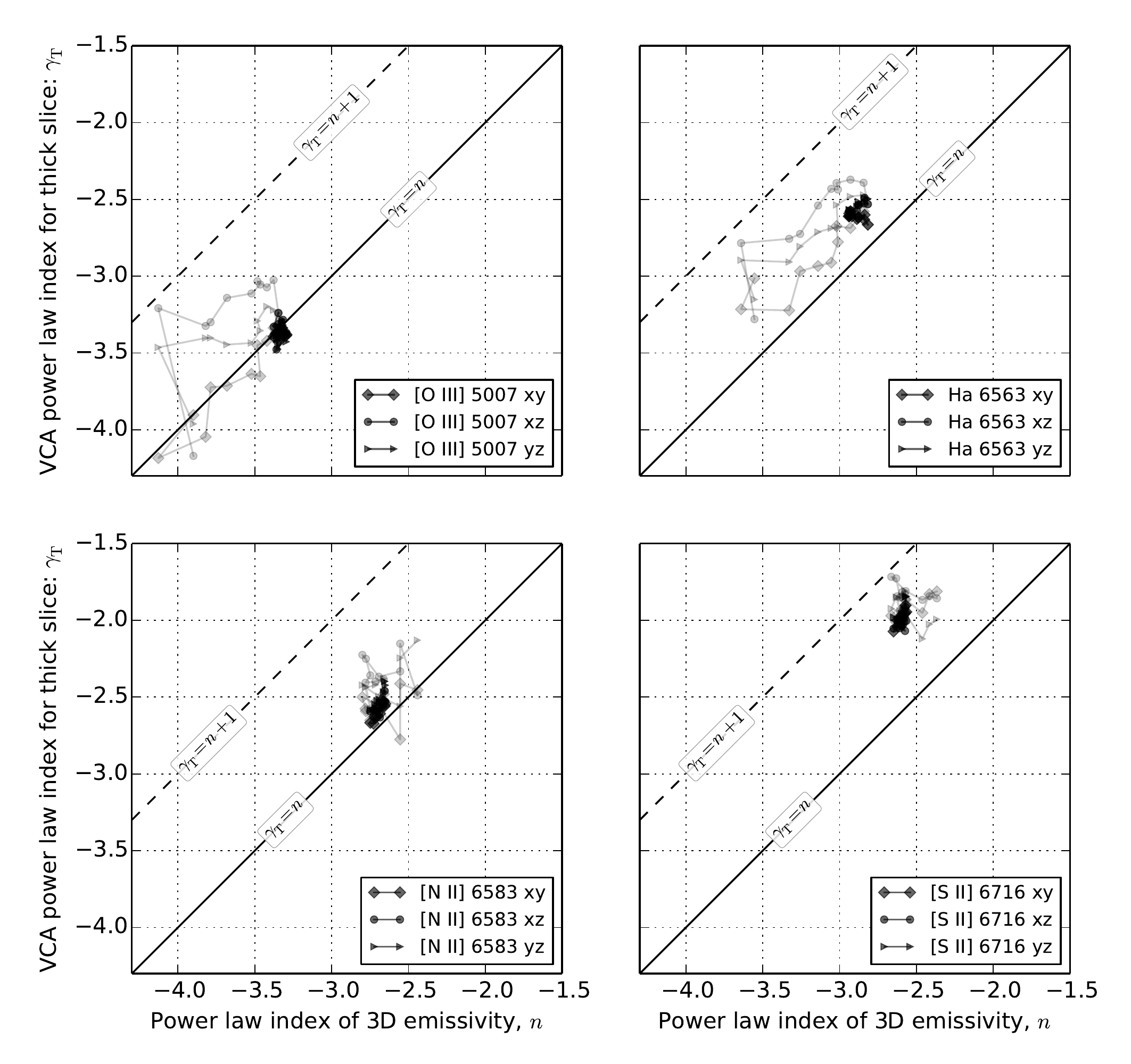}
  \caption{As Fig.~\ref{fig:sf-vs-n}, but showing VCA power spectrum
    slopes for thick slices, \gammaVCAvthick{}, versus slope, \(n\),
    of the intrinsic emissivity power spectrum for each emission line.
    The theoretical expectations are shown by diagonal lines for the
    cases where the line-of-sight depth of the emitting region is
    larger (continuous line) or smaller (dashed line) than the
    transverse scales that are sampled.  }
  \label{fig:vca-thick-vs-n}
\end{figure*}

The slopes of the power spectra of the thick slices themselves, which
are simply the velocity-integrated surface brightness images\footnote{
  Although for simplicity, extinction is not included.}  are predicted
\citep {2000ApJ...537..720L} to be equal to the slopes of the 3D power
spectra of their respective emissivities.  The comparison between
these two quantities is shown in Figure~\ref{fig:vca-thick-vs-n}, from
which it is clear that only in the case of \oiii{} are the two slopes
equal.  In the case of the other lines, \gammaVCAvthick{}
is shallower than the emissivity's spectral index \(n\) by 0.36, 0.19,
0.61 for \ha, \nii, and \sii, respectively.  The reason for this
discrepancy is the increasingly ``sheet-like'' morphology of the
emission in the lower ionization lines.  As shown in \S~4.1 of \citet
{2003ApJ...593..831M}, one should see a transition from
\(\gammaVCAvthick{} = n\) to the shallower slope
\(\gammaVCAvthick{} = n + 1\) at transverse scales larger than
the line-of-sight depth of the emitting region.

\subsection{Comparison with observational results}

Non-thermal linewidths in \hii{} regions have been interpreted as
evidence for turbulence in the photoionized gas. Galactic \hii{} regions
generally exhibit subsonic turbulent widths
\citep{{1985ApJ...288..142R},{1986ApJ...304..767O}}, while giant
extragalactic \hii{} regions can have supersonic turbulent motions
\citep{{2004Ap&SS.292..111B},{2011MNRAS.413..705L}}.  The majority of
observational studies consider only the H$\alpha$ line, but
\citet{1988AA...198..283O} and \citet{{1988ApJS...67...93C}} examine
the components of the \oiii$\lambda$5007 line in the Orion Nebula,
M42, while \citet{1992ApJ...387..229O} and \citet{1993ApJ...409..262W}
investigate the kinematics of the [S{\sc iii}] and [O{\sc i}] lines of
this same object.  The giant extragalactic \hii{} region NGC-595 was
studied by \citet{{2011MNRAS.413..721L}} who analyzed the H$\alpha$,
\oiii{} and \sii{} kinematics.  The heavier ions have the advantage that
their emission lines have low thermal broadening compared to the
H$\alpha$ line.

Observational studies of the spatial scales of velocity fluctuations
have mostly focused on the structure function of velocity centroids.
The results are rather disparate, partly because the methodology
varies considerably between different studies.  For instance, some
authors attempt to filter out ``ordered'' large-scale motions before
analysing the fluctuations \citep{1995ApJ...454..316M,
  2011MNRAS.413..721L}, whereas others analyse the unfiltered
observations \citep{1992ApJ...387..229O, 1997ApJ...487..163M}.  Also,
in some cases multiple Gaussian velocity components are fitted to the
line profiles \citep{1988ApJS...67...93C, 1993ApJ...409..262W}, which
are then assigned to a small number of velocity ``systems'' that are
each analysed separately, whereas in most studies the mean velocity of
the entire line profile is used.

Despite these differences, there are interesting commonalities in the
results: a rising structure function with \(m_{\mathrm{2D}} =
0.5\)--\(1.0\) is nearly always found at the smallest scales, which
transitions to a flat structure function with \(m_{\mathrm{2D}} \sim
0\) at larger scales.  However, the scale at which the transition
occurs varies enormously from object to object, from
\(0.02\)--\(0.2\)~pc in compact ($R = 1$ to 5~pc) Galactic \hii{}
regions \citep{1987ApJ...317..676O, 1988ApJS...67...93C,
  1993ApJ...409..262W, 1995ApJ...454..316M}, up to 50~pc in giant (\(R
\sim 400\)~pc) extragalactic regions \citep{2011MNRAS.413..721L}. We
comment that the sound-crossing time for a region of size 50~pc is
about 5~Myr, roughly the same as the estimated age of the NGC~595
nebula \citep{1990ApJ...364..496D}. For the full extent of NGC~595,
the sound crossing time is about 40~Myr. It is therefore unlikely that
the turbulence in such large regions is in a statistically steady
state unless it is highly supersonic. Indeed,
\citet{1988AA...198..283O} suggest that in the case of large
extragalactic \hii{} regions, the large linewidths could be due to
multiple velocity components, that is, parts of the \hii{} region
with separation greater than the distance a sound wave could travel
within  the current lifetime of the object
are kinematically
distinct.  Such giant \hii{} regions show velocity centroid
dispersions of \(\sigma_{\mathrm{c}} > 10~\mathrm{km\ s^{-1}}\) on the
largest scales, which is several times larger than is seen in compact
single-star regions or in our simulations.  We will therefore not
consider them further since they are governed by additional physical
processes, such as powerful stellar winds and the cluster
gravitational potential, which are beyond the scope of the current
paper.

The explanations that have been offered for the break in the structure
function slope are also varied.  In the case of compact \hii{}
regions, it is often taken to indicate the characteristic
line-of-sight depth of the emission zone \citep{1951ZA.....30...17V,
  1987ApJ...317..686O}, with projection smoothing steepening the slope
at the smaller separations (see \S~\ref{subsec:projsmooth} above).
If that were the case, then the correct three-dimensional structure
function slope is the flat one: \(m_{\mathrm{3D}} \sim 0\),
corresponding to a velocity power spectrum slope of \(n = -3\).  This
interpretation would be broadly consistent with our simulation
results, which show a very similar velocity power spectrum
(Fig.~\ref{fig:ps}).  However, our simulated structure functions
rarely show a clear break in the same way as the observations do,
although they do show a slight negative curvature in many cases.  This
is probably because of the very limited useful dynamic range, roughly
a factor of 10, that the simulations allow between the small scales
that are affected by numerical diffusion and the large scales, that
are affected by systematic flows, anisotropies, and edge-effects.

An alternative explanation for the observed break in the structure function is that
it represents the scale of the largest turbulent eddies
\citep{1988ApJS...67...93C, 1995ApJ...454..316M}
and that the fluctuations at larger scales are simply uncorrelated.  
In such a picture it would still be necessary to postulate a velocity spectrum
considerably shallower than Kolmogorov in order to explain the small-scale slope.

Based on the discussion of our simulation results above
(\S~\ref{sssec:vca2}), it seems that Velocity Channel Analysis
would be a very useful complement to the structure function, since it
is less affected by uncertainties in projection smoothing and gives a
more consistent result between different emission lines (at least, for
our simulations).

In a forthcoming paper, we will present such an analysis of recent
high-resolution echelle spectroscopy of multiple emission lines in the Orion Nebula
\citep{2008RMxAA..44..181G, 2008AJ....136.1566O}.

\section{Summary}
\label{sec:summary}
\begin{enumerate}
\item We have investigated the statistics of fluctuations in physical
  conditions within a radiation-hydrodynamic simulation of the
  evolution of an \hii{} region inside a highly inhomogeneous
  molecular cloud.  We find that steady-state turbulence,
  corresponding to time-independent profiles of the 3D power spectra, is
  only established after about 1.5 sound-crossing times of the \hii{}
  region. In these simulations, this corresponds to about 200,000
  years (\S~\ref{sssec:pspec}).
\item We find a power-law behaviour for the 3D power spectra in the
  range from about 1 pc down to 0.125 pc, equivalent to 16
  computational cells. The larger scale can be interpreted as the size
  of the largest photoevaporated flows, while the smaller scale is
  about twice the numerical dissipation scale.  The power-spectrum
  slopes of the velocity and density fluctuations are very similar and
  always lie in the range \(-3.1 \pm 0.1\).  This is significantly
  shallower than the slope predicted for the inertial range of either
  incompressible or compressible turbulence ($-3.667$ to $-4.1$).
  This suggests that turbulent driving is occuring over all scales in
  our simulation, unlike the case of classical turbulence where energy is
  injected only at the largest scales.  The power-spectrum slopes of
  the emissivities of optical lines are even shallower, increasingly so
  for lower ionization lines, indicating that the smallest scale
  fluctuations are dominant (\S~\ref{sssec:ips}).
\item We investigate in detail the utility of observational
  diagnostics for inferring the power spectra of emissivity and
  velocity fluctuations in our simulation.  We find that the
  traditional velocity centroid structure function technique gives
  ambivalent results because of the effects of projection smoothing,
  combined with the fact that the effective line-of-sight depth of the
  emitting gas does not have a single well-defined value.  In
  addition, the presence of anisotropic motions such as champagne
  flows can yield misleading structure function slopes when the
  simulation is viewed from certain directions (\S~\ref{sssec:strfunc}).
\item The more recently developed technique of Velocity Channel
  Analysis is found to offer a more robust diagnostic of the
  three-dimensional velocity statistics of our simulation.  The slope
  of the velocity power spectrum can be correctly recovered to a precision
  of \(\pm 0.1\) from either high or low ionization lines, and with no
  significant dependence on viewing direction (\S~\ref{sssec:vca2}).
\end{enumerate}

\section*{Acknowledgments}
We would like to thank the referee for constructive comments, which
improved the presentation of this paper.  SNXM acknowledges a CONACyT,
Mexico student fellowship. SJA would like to thank DGAPA-UNAM for
financial support through project IN101713. SJA, WJH and GM thank
Nordita for support during the program Photo-Evaporation in
Astrophysical Systems. This work has made use of NASA's Astrophysics
Data System.

\appendix

\section[]{Example second-order structure functions of the line-of-sight velocity centroids}
\label{app:sf}
\begin{figure*}
  \centering
  \includegraphics[width=\textwidth]{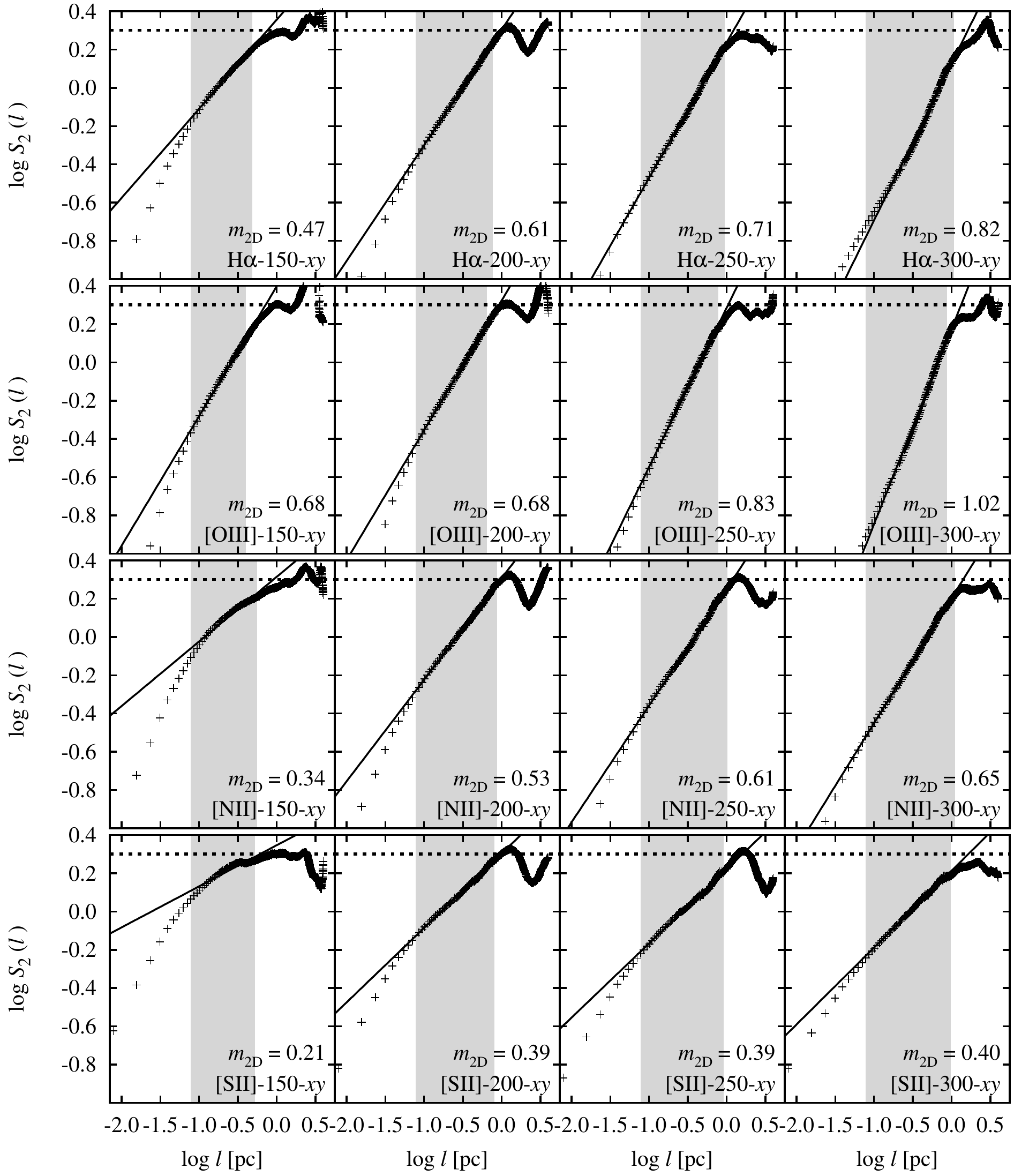}
  \caption{Second-order structure functions against length scale for
    projection onto the $xy$-plane. From top to bottom: H$\alpha$,
    \oiii$\lambda 5007$, \nii$\lambda 6584$, \sii$\lambda 6716$. From
    left to right: 150,000, 200,000, 250,000 and 300,000~years. The
    points represent the calculated structure function for the
    numerical simulation. The solid line is the least-squares fit to
    the data points between limits described in the text, represented
    by the grey rectangle. The horizontal dotted line at $\log 2$ is
    included as a reference value. The index $m_\mathrm{2D}$ of each power-law fit
   is indicated in the corresponding panel.}
\label{fig:sfunc}
\end{figure*}

\begin{figure*}
 \centering
 \includegraphics[width=\textwidth]{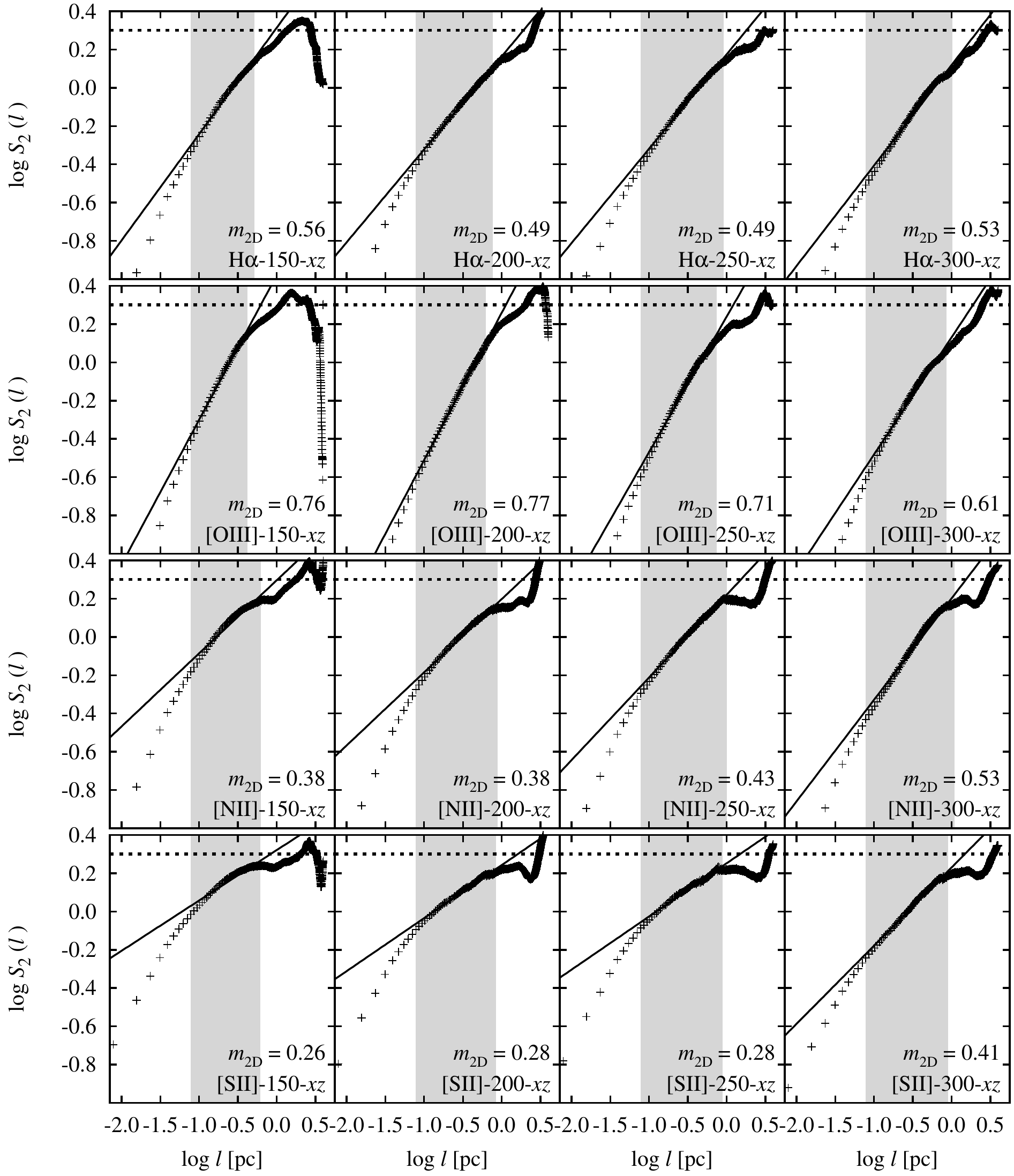}
 \caption{Same as Fig.~\protect\ref{fig:sfunc} but for a projection
   onto the $xz$ plane.}
 \label{fig:sfuncxz}
\end{figure*}

\begin{figure*}
  \centering
  \includegraphics[width=\textwidth]{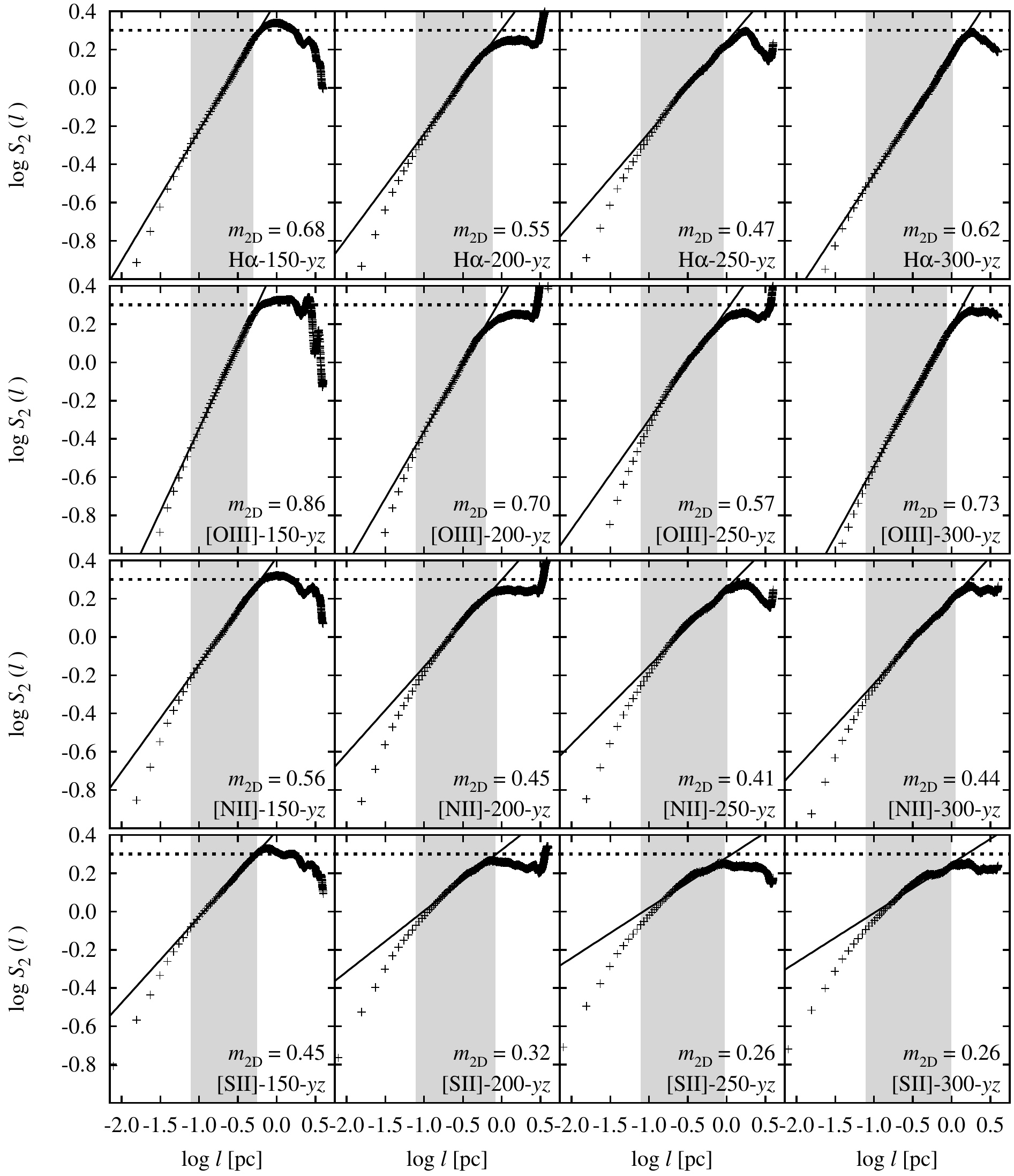}
  \caption{Same as Fig.~\protect\ref{fig:sfunc} but for a projection
    onto the $yz$-plane.}
  \label{fig:sfuncyz}
\end{figure*}

Figures~\ref{fig:sfunc} to \ref{fig:sfuncyz} show the second-order
structure functions of the line-of-sight velocity centroid maps (see
\S\S~\ref{sssec:strfunc} and \ref{subsubsec:centroid}) for the four
emission lines at the four evolutionary times depicted in
Figure~\ref{fig:HIIimages}.  If turbulence is present, the
second-order structure function should exhibit an inertial range over
which it is a power law with length scale. Accordingly, we perform a
least-squares fit to the data points. However, it is not immediately
clear what the limits for the fit should be. 

At small scales, the lower limit for the inertial range should be
defined by the scale at which numerical dissipation effects cease to
be important \citep {2004ApJ...604..196B}. For the present
simulations, we tested several values and the size scale equivalent to
10 computational cells proved to be adequate for all emission lines
and evolution times studied. For the upper limit, we examined the
projected emission maps and calculated the area occupied by the pixels
having greater than the mean intensity. This method is independent of
the resolution of the image and could equally be applied to images
obtained from observations. We then took the radius of the circle
having the same area to be the upper limit for the least-squares
fit. This procedure appears to work very well, as can be seen in
Figures~\ref{fig:sfunc} and \ref{fig:sfuncyz}. If a different line of
sight is chosen, the radius of this circle will be different and needs
to be calculated self-consistently for every projection.  Note that
the inertial range for each combination of line and view tends to
become broader with time due to the expansion of the \hii{} region.
At the latest time, 300,000~yrs, both the H$\alpha$ and \oiii$\lambda
5007$ structure functions appear to develop a break, which would be
better fit by two power laws, one below a scale of about 0.3~pc and a
steeper one for larger scales. However, we have fit just a single
power law to both of these cases. We speculate that this apparent
break in the power law at late times could be due to energy injection
at size scales associated with the photoevaporated flows emanating
from the ubiquitous clumps and filaments seen in the emission-line
images (see Fig.~\ref{fig:HIIimages}).

\begin{figure}
  \centering
  \includegraphics[width=0.8\linewidth]{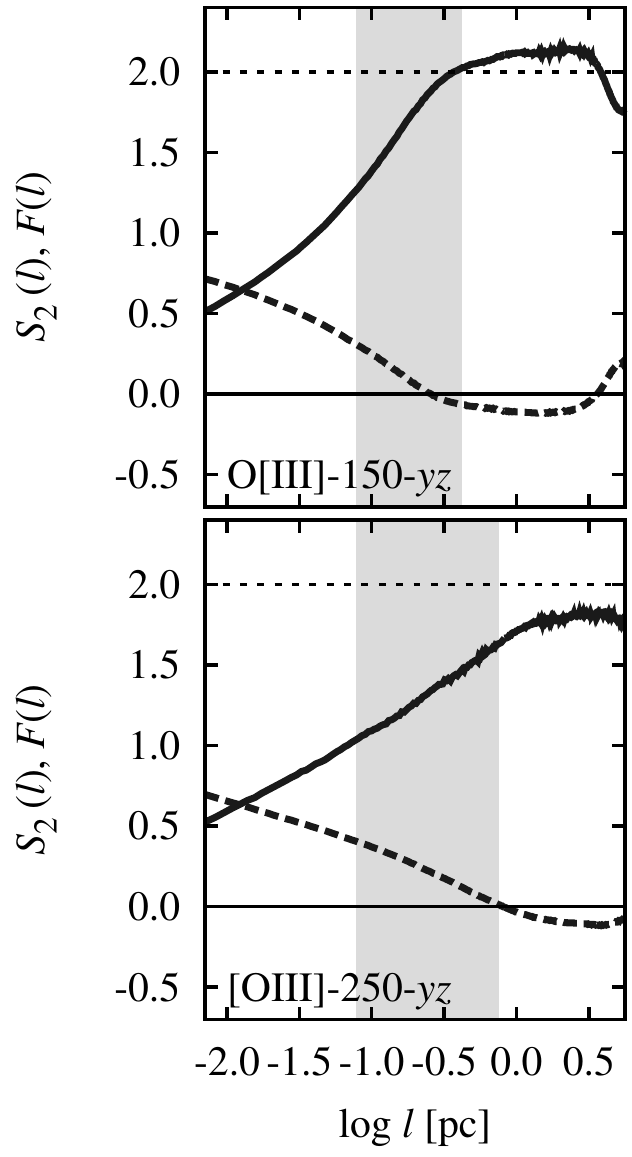}
  \caption{Second-order structure function (thick solid line) and auto-correlation
    function (thick dashed line) against logarithmic length scale. The
    horizontal lines at 2 and 0 are included as reference values.}
\label{fig:sfauto}
\end{figure}

An alternative criterion for the upper limit was used by \citet
{2011MNRAS.413..721L} who used the theoretical result for isotropic,
homogeneous turbulence that decorrelation of the second-order
structure function occurs when the auto-correlation function changes
sign from positive to negative. In the theory, this corresponds to the scale for
which the second-order structure function is equal to
2. Figure~\ref{fig:sfauto} shows the structure function and
auto-correlation function obtained from our simulations at times 150,000~yrs and 250,000~yrs for the
\oiii$\lambda 5007$ emission line velocity centroids projected onto
the $yz$-plane. The other emission lines and viewing directions give
very similar results. From this figure, we see that at early times,
when the structure function does indeed rise above 2, this corresponds
approximately to the length scale at which the auto-correlation
function changes sign. However, at later times the auto-correlation
function changes sign at a length scale much smaller than the size of
the computational box but the structure function remains less than
2. This suggests that the assumptions of the simple theory (isotropic,
homogeneous turbulence) do not apply to the emission-line velocity
centroids obtained from our simulations. Figure~\ref{fig:sfauto} also
indicates the fitting range suggested by the procedure described
earlier. At early times, our procedure results in a slightly larger
upper length scale for the power-law fit than using the zero point of
the auto-correlation function. At later times, the two more-or-less
coincide and this is true for all the emission lines, viewing angles
and times we have examined. Since there is no clear reason to prefer
the auto-correlation approach, we will use our computationally simpler
procedure to determine the upper length scale for the power-law fit.


\section[]{Example power spectra from Velocity Channel Analysis}
\label{app:vca}
\begin{figure*}
\centering
\includegraphics[width=\textwidth]{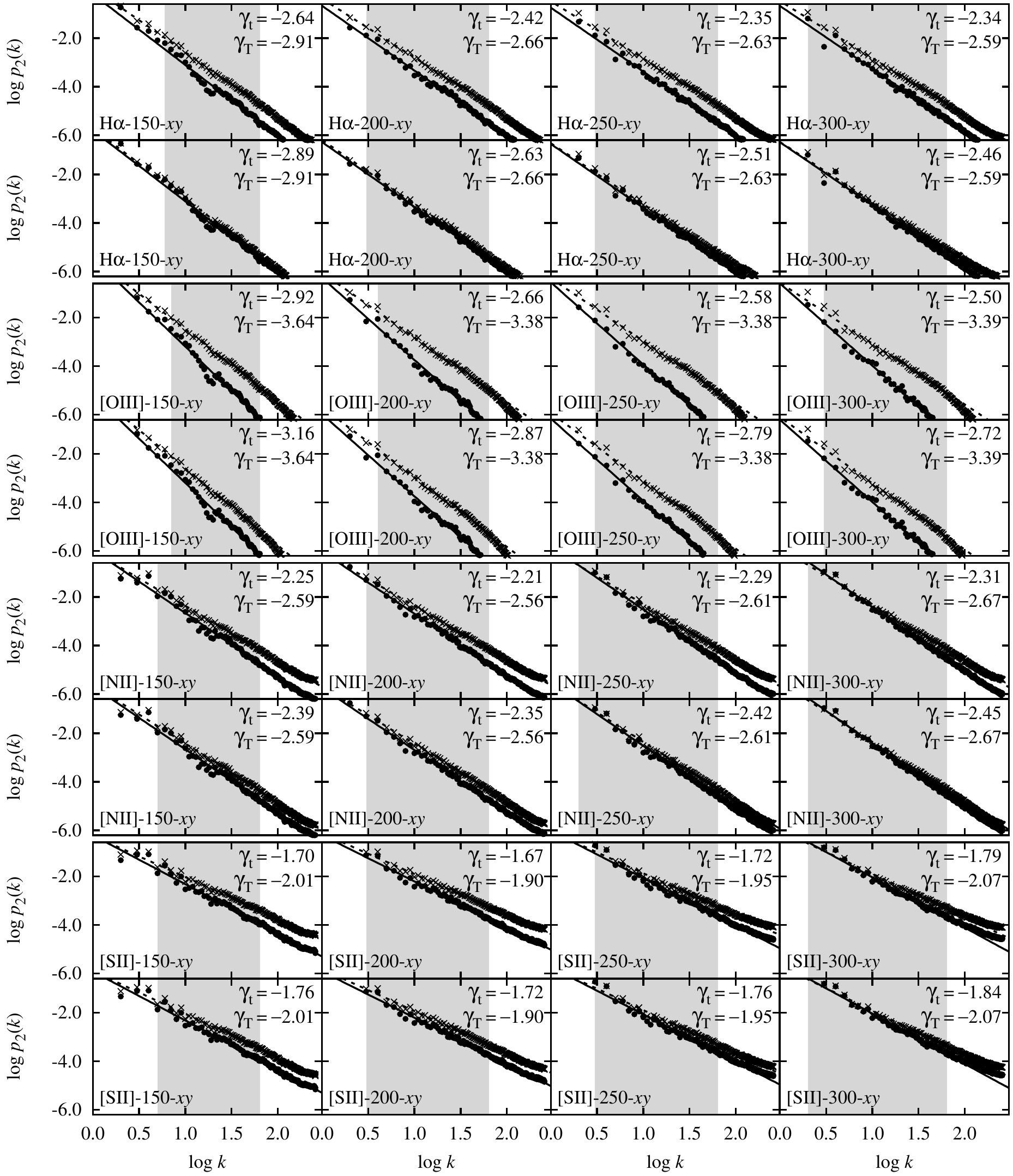}
\caption{Power spectra of the velocity channels. From top to bottom:
  H$\alpha$, \oiii$\lambda 5007$, \nii$\lambda 6584$, \sii$\lambda
  6716$. From left to right: 150,000, 200,000, 250,000 and
  300,000~years. The upper panel in each figure are for velocities
  without thermal broadening, the lower panel is the case with thermal
  broadening. The points represent the calculated power spectra for
  the numerical simulation: Crosses--thin slice ($n=32$ channels),
  filled circles--thick slice
  ($n=1$ channel). The dashed line is the
  least-squares fit to the data points for the thin slice between limits described in the
text represented by the grey rectangle. The solid line is the least-squares fit to
the data points for the thick slice. The indices \gammaVCAthin{} (thin
slice) and \gammaVCAvthick{} (thick slice) of each power-law fit
   are indicated in the corresponding panels.}
\label{fig:vca}
\end{figure*}

\begin{figure*}
 \centering
 \includegraphics[width=\textwidth]{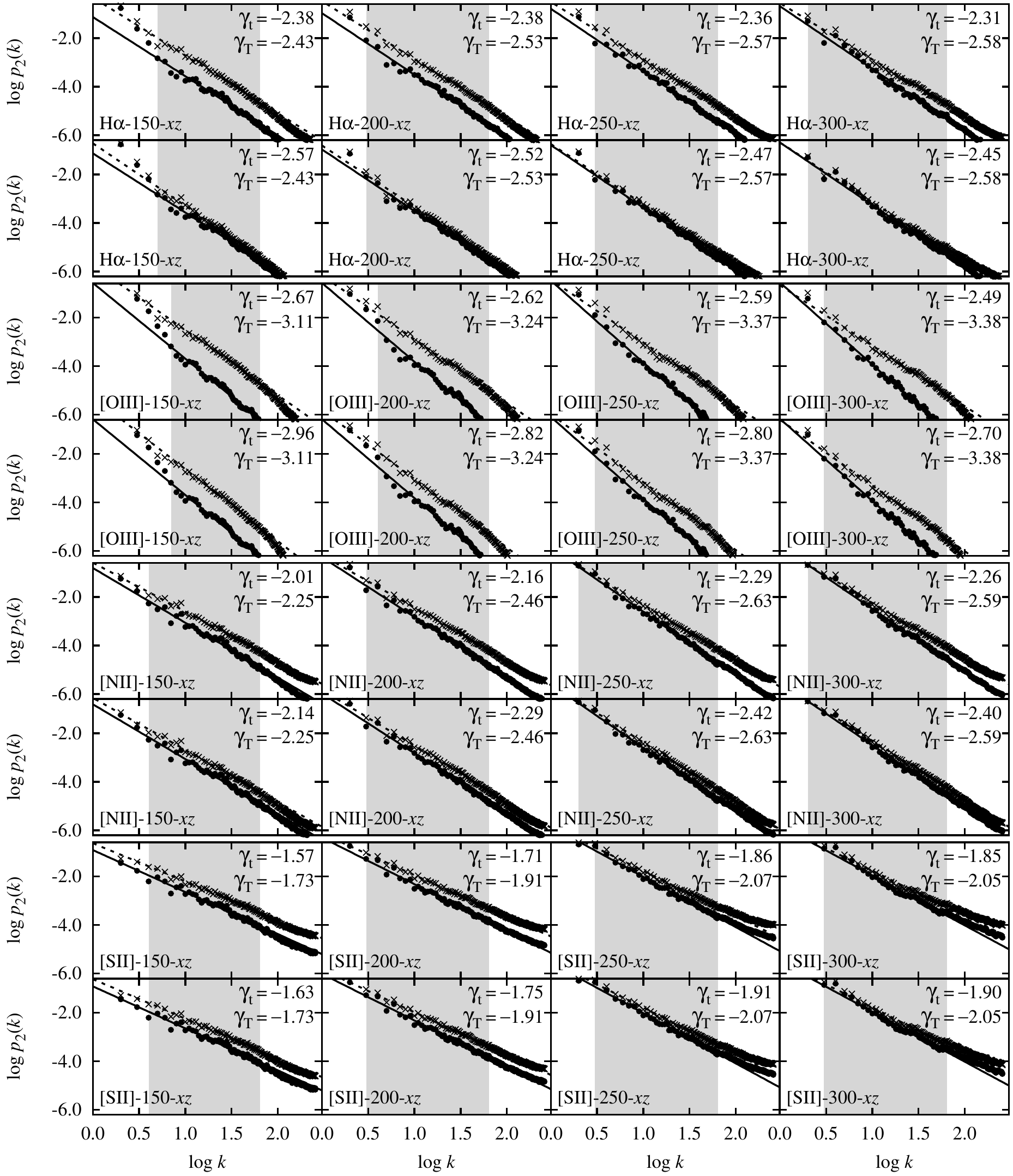}
 \caption{Same as Fig.~\protect\ref{fig:vca} but for a projection onto the $xz$ plane.}
 \label{fig:vcaxz}
 \end{figure*}

\begin{figure*}
 \centering
 \includegraphics[width=\textwidth]{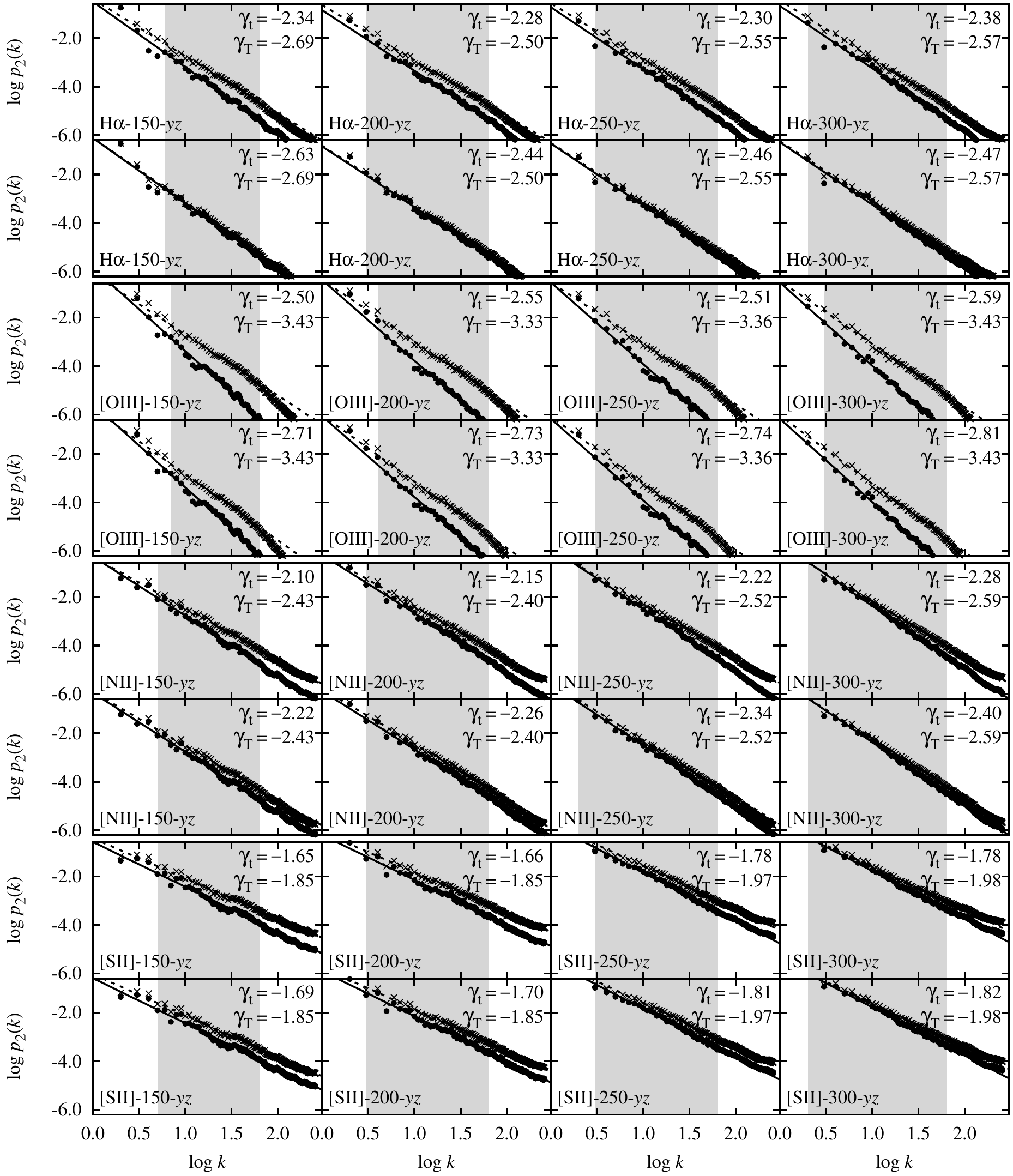}
 \caption{Same as Fig.~\protect\ref{fig:vca} but for a projection onto the $yz$-plane.}
 \label{fig:vcayz}
 \end{figure*}

Figures~\ref{fig:vca} to~\ref{fig:vcayz} show the power spectra
resulting from the velocity channel analysis (see
\S~\ref{sec:stats-vca}). Each of the three figures is for a different
viewing direction and shows the four emission lines at four different
times. For each combination of line and time, there are two panels: an
upper panel without including thermal Doppler broadening and a lower
panel with the broadening effects included.  In each graph, two power
spectra are plotted: one representing a very thick velocity slice
(i.e., encompassing all the emission) and the other averaged over thin
velocity slices of width $\delta v \sim 5$~km~s$^{-1}$.  Also shown
on each panel are the power-law indices obtained from
least-squares power-law fits to the thin (\gammaVCAthin{}) and very
thick slice (\gammaVCAvthick{})
spectra and the range in wavenumber over which the fit is
calculated. This wavenumber range corresponds to the length-scale
range used for the structure function fits (see
\S~\ref{sssec:s2func}).  The very thick velocity slice is equivalent
to the total intensity along the line of sight and its power spectrum
does not vary with the addition of thermal broadening.

It is clear that the thermal broadening has a large effect on the VCA
of the H$\alpha$ line, effectively erasing the difference in slope
between the thin and thick slices.  For photoionized gas at $T_e=
10^4$~K, the FWHM of the H$\alpha$ line is $\sim 22$~km~s$^{-1}$,
while that of an oxygen line is a quarter of this, $\sim
5.5$~km~s$^{-1}$.  Indeed, the heavier ions are less affected by
thermal broadening, but a slight steepening of the thin-slice power
spectra can still be seen, amounting to a reduction in \gammaVCAthin{}
of \(\sim 0.1\).

For the thermally broadened case, the variation with time of the
slopes of these fits, \gammaVCAvthick{} for the thick slices and
\gammaVCAthin{} for the thin slices, is shown in
Fig~\ref{fig:vcatrends} and discussed in \S~\ref{sssec:vca}.


\label{lastpage}

\end{document}